# A Map of the Zintl $AM_2Pn_2$ Compounds: Influence of Chemistry on Stability and Electronic Structure

Andrew Pike[1], Zhenkun Yuan[1], Gideon Kassa[1], Muhammad R Hasan[2], Smitakshi Goswami[1,3], Sita Dugu[4], Shaham Quadir[4], Andriy Zakutayev[4], Sage Bauers[4], Kirill Kovnir[2,5], Jifeng Liu[1], Geoffroy Hautier[1]

1 Thayer School of Engineering, Dartmouth College, Hanover, NH 03755, USA

2 Department of Chemistry, Iowa State University, Ames, IA 50011, USA

3 Department of Physics and Astronomy, Dartmouth College, Hanover, NH 03755, USA

4 National Renewable Energy Laboratory, Golden, CO 80401, USA

5 Ames National Laboratory, U.S. Department of Energy, Ames, IA 50011, USA

**Abstract**

The $AM_2Pn_2$ (A= Ca, Sr, Ba, Yb, Mg; M= Mn, Zn, Cd, Mg; and Pn=N, P, As, Sb, Bi) family of Zintl phases has been known as thermoelectric materials and has recently gained much attention for highly promising materials for solar absorbers in single junction and tandem solar cells. In this paper we will, from first-principles, explore the entire family of $AM_2Pn_2$ compounds in terms of their ground state structure, thermodynamic stability, and electronic structure. We also perform photoluminescence spectroscopy on bulk powder and thin film samples to verify our results, including the first measurements of the bandgaps of $SrCd_2P_2$ and $CaCd_2P_2$. The $AM_2Pn_2$ compounds exhibit broad stability, are mostly isostructural in the $CaAl_2Si_2$-type structure ($P\bar{3}m1$), and cover a wide range of bandgaps from 0 to beyond 3 eV. This could make them useful for a variety of purposes, for which we propose several candidates, such as $CaZn_2N_2$ for tandem top cell solar absorbers and $SrCd_2Sb_2$ and $CaZn_2Sb_2$ for infrared detectors. By examining the band structures of the $AM_2Pn_2$, we find that $Mg_3Sb_2$ has the most promise as a thermoelectric material due to several off-Γ valence band pockets which are unique to it among the compositions studied here.

## Introduction

Zintl compounds are a vast category of inorganic compounds that exhibit a combination of ionic and covalent bonding. Cations donate electrons, allowing a covalently bonded polyanion to form[11–3]. Zintl compounds have a valence-precise composition, which allows for interchangeability of ions with like charges forming new compounds similar to the parent. Examples of Zintl compound families include $A_{11}MPn_{14}$, $A_{21}M_4Pn_{18}$, $A_9M_4Pn_{14}$, and $AM_2Pn_2$, among many others, where the A cation is typically an alkaline earth metal, the M cation is a main group or transition metal, and Pn is a pnictogen[4]. Zintl compounds are typically semiconductors because of the ionic nature of the interactions between cations and polyanions. Due to the narrow bandgap of antimonides and bismuthides, they have mostly been investigated for their thermoelectric properties. The Zintl concept allows for straightforward tunability of the electronic structure, often coupled with complex unit cells, thus offering a unique platform to tune the electrical properties with a low lattice thermal conductivity.

The $AM_2Pn_2$ compounds, sometimes called the "1-2-2" compounds, are a compositionally diverse class of materials. We will restrict our investigation and discussion of the $AM_2Pn_2$ to A= Ca, Sr, Ba, Yb, Mg; M= Mn, Zn, Cd, Mg; and Pn=N, P, As, Sb, Bi. Many of the original syntheses of the $AM_2Pn_2$ compounds were performed beginning in the late 1970s through the 1980's[5–13]. From those syntheses, it was revealed that most of these compounds are isostructural, crystallizing in the $CaAl_2Si_2$-type structure (space group $P\bar{3}m1$, No. 164) shown in Figure 1, and do not show any sensitivities to air or moisture exposure. In the past 20 years, the $AM_2Pn_2$ compounds crystallizing in the $CaAl_2Si_2$-type structure with small to intermediate bandgaps have attracted attention for potential applications in heat-to-electricity conversion as thermoelectric materials, such as in systems based on $YbZn_2Sb_2$ [14–16] and $Mg_3Sb_2$ [17–20]. Very recently, it was discovered via high-throughput computational screening that $BaCd_2P_2$ and $CaZn_2P_2$ are promising thin-film solar absorbers with high optical absorption in the visible light range and favorable defect properties allowing for long carrier lifetimes [21,22]. Driven by these successes we are motivated to search the entire family of $AM_2Pn_2$ compounds for other promising material candidates for thermoelectric devices, photovoltaic solar cells, as well as infrared detectors.

Here, we have systematically studied the phase stability and electronic band structure of the $AM_2Pn_2$ compounds in a wide range of compositions, using first-principles calculations finding the majority of compositions studied here are stable and isostructural and that these possess a wide range of bandgaps. Bandgap generally increases as Pn mass decreases; while bismuthides are mostly metallic, certain nitrides have bandgaps greater than 3 eV. We compare the results to experimental literature and complement with new experimental results. Our work can be used to understand well-established $AM_2Pn_2$ compounds, suggest new $AM_2Pn_2$ compounds to synthesize as well as to suggest specific applications for certain chemistries.

## Methodology

Our first-principles calculations were performed using the projector augmented-wave (PAW) pseudopotential method as implemented in the VASP code [23–27]. All the calculations were automated using the Atomate code, including generation of VASP input files [28]. To determine the equilibrium crystal structure for each composition, we considered each $AM_2Pn_2$ adopting three different structures, shown in Figure 1, and these were first relaxed using the PBE functional, which became the starting point for further calculations [29]. For the PBE structure relaxations, an energy cutoff of 520 eV was used for the plane wave basis set and other settings including **k**-point density and pseudopotentials are consistent with the Material Project database [30]. For Yb the f-electrons were considered frozen in the core.

Next, we performed a further full geometry optimization, using the revised and renormalized strongly constrained and appropriately normed ($r^2$SCAN) functional [31,32]. For the $r^2$SCAN calculations, an increased plane wave energy cutoff of 680 eV was used, the smallest distance allowed between **k**-points was 0.22 Å$^{-3}$, and the PBE_54 set of pseudopotentials was used. From our preliminary tests, PBE fails to predict that several experimentally reported $AM_2Pn_2$ compounds are stable, whereas $r^2$SCAN calculations agree with previous experimental results (see SI Section 2 for a further discussion of PBE and $r^2$SCAN results). Specifically, this error occurs for the Mn-containing $AM_2Pn_2$ compounds. The ground-state structure was determined by comparing the calculated total energies of the different polymorphs tested.

To determine the 0K phase stability, competition with other elemental, binary, and ternary phases must be considered. To do this using the $r^2$SCAN functional, the decomposition products of each $AM_2Pn_2$ were queried from the Materials Project (MP, which uses primarily PBE-GGA for its stability calculations) and relaxed in $r^2$SCAN to be considered in the following reaction:

$$MP\ Decomposition \rightarrow AM_2Pn_2 \quad (1)$$

This prevents the calculation of the entire phase diagram, reducing computational cost, but gives a realistic reference for the phase stability, rather than a formation energy calculated relative to the energy of the pure elemental phases. From the calculated total energies, the decomposition energy can be calculated to allow for assessment of the stability of the $AM_2Pn_2$.

$$E_{hul} \ = E_{AM_2Pn_2} - \Sigma n * E_{products} \quad (2)$$

Where $n$ is the reaction coefficient and $E_{products}$ and $E_{AM_2Pn_2}$are the total energies of the decomposition products and the $AM_2Pn_2$ compound of interest, respectively. Eq. (1) represents the chemical reaction studied in order the calculate the $E_{hull}$ in Eq. (2). Here, negative $E_{hul}$ indicates a phase is thermodynamically stable. Taking $BaCd_2As_2$ as an example, the Materials Project predicts that it lies in a region of the Ba-Cd-As phase diagram that is bounded by $Ba_2Cd_2As_3$, $Cd_3As_2$, and Cd, so for its decomposition we consider the reaction:

$$0.5\ Ba_2Cd_2As_3 \ + \ 0.25\ Cd_3As_2 \ + \ 0.25\ Cd \rightarrow BaCd_2As_2 \quad (3)$$

So, its $E_{hull}$ would be

$$E_{hull,BaCd_2As_2} = E_{BaCd_2As_2} - 0.5\ E_{Ba_2Cd_2As_3} \ - \ 0.25\ E_{Cd_3As_2} - \ 0.25E_{Cd} \quad (4)$$

For clarity we have used to the term $E_{hul}$ here since it is well understood from other similar investigations, but we note that for the sake of this paper, we have defined it slightly differently than the common definition. Usually the $E_{hull}$ is calculated from the decomposition of the compound being investigated based on its full phase diagram, but we did not explicitly compute the entire phase diagram to get the full hull. For computational efficiency we have only calculated with $r^2$SCAN a limited set of compounds for each chemical system, relying on the Materials Project prediction of the decomposition, which is done at the PBE level of theory. In practice this should be identical to the $E_{hull}$ calculated from $r^2$SCAN, except for the case where the set of compounds on the hull changes when calculated in $r^2$SCAN compared to those reported on the Materials Project.

To determine the electronic band structure, non-collinear hybrid functional calculations using the parameterization of Heyd, Scuseria, and Ernzerhof (HSE06) including spin-orbit coupling (SOC) were performed based on the PBE-relaxed structures using a plane wave energy cutoff of 520 eV, the PBE_54 set of pseudopotentials, and the HSE mixing parameter (α) was set to 0.25 [33,34]. Computing HSE band structures on PBE relaxed structures saves a significant amount of computational time compared to HSE band structure calculations on structures also relaxed from HSE. In this way, the calculations balance accuracy and computational cost. We have computed the carrier effective masses, without including SOC. We do not expect significant differences between an effective mass calculated from an HSE and HSE+SOC band structure, since the difference in functional should not significantly affect the shape of the bands near the bandgap. The conductivity effective mass tensor was computed using Boltztrap2. This conductivity effective mass effectively takes into account the effects of band non-parabolicity and multiple bands' contributions to transport. The effective mass tensors were calculated assuming a temperature of 300 K [35]. We report an average of the three principal directions of transport (trace of the tensor divided by three). We do not report the conductivity effective mass for materials with a bandgap less than 0.1 eV[36].

**Bulk samples synthesis and XRD characterization.**

Powder synthesis of $BaCd_2P_2$, $CaCd_2P_2$, $SrCd_2P_2$, $BaCd_2As_2$, and $SrCd_2Bi_2$ were attempted using elements in the stoichiometric 1:2:2 ratio (A:M:Pn). See SI Section 1 for safety information for performing these syntheses. Elemental Ba (99.9%, Alfa Aesar), Ca (99.98%, Alfa Aesar), Sr (99.90%, Sigma Aldrich), Cd (99.95%, Alfa Aesar), red phosphorus (98.90%, Alfa Aesar), As (99.999995%, Furukawa Denshi Co.), and Bi (99.998%, Alfa Aesar) were used for their respective reactions. The reaction mixtures were placed into carbonized 9/11 mm inner/outer diameter silica ampoules under Ar atmosphere in a glovebox. The ampoules were evacuated and then sealed using a hydrogen-oxygen torch. The sealed ampoules were heated in a muffle furnace to a temperature ramp up to 800 $^oC$-1000 $^oC$ over 8-10 hours and annealed at that temperature for 48 hours. After cooling in the turned-off furnace, the ampoules were then opened in ambient conditions. The resulting samples were ground in agate mortars and reannealed at 800 $^oC$-1000 $^oC$ for 48 hours in an evacuated carbonized silica ampoule. The synthesized powder samples were characterized by powder X-ray diffraction (PXRD). PXRD patterns were collected using a Rigaku MiniFlex 600 benchtop diffractometer with Cu-$K_\alpha$ radiation ($\lambda$ = 1.5406 Å) and Ni-$K_\beta$ filter in $3° \leq 2\theta \leq 90°$ range with 0.02° steps at 10° /min. Four samples, $BaCd_2P_2$, $CaCd_2P_2$, $SrCd_2P_2$, and $BaCd_2As_2$ which were single-phase according to powder X-ray diffraction, were used for subsequent analyses. The $SrCd_2Bi_2$ sample has no target phase but rather a mixture of $SrCdBi_2$ + Cd. See Fig. S5 for PXRD patterns of the compounds investigated here.

**Thin film samples synthesis and XRD characterization.**

Thin films of $CaZn_2P_2$ and $SrZn_2P_2$ were synthesized by radio frequency (RF) co-sputtering with 50.8 mm diameter metallic Ca, Sr and Zn targets. The RF power densities applied to the targets were maintained in the range of 1.48 – 1.97 W cm$^{-2}$ for Ca, 1.13 – 1.25 W cm$^{-2}$ for Sr, and 2.96 W cm$^{-2}$ for Zn. Base pressure of the growth chamber was maintained below $10^{-7}$ Torr prior to deposition. The deposition was conducted in 2% $PH_3$/98% Ar gas mixture at a flow rate of 19.5 sccm, with the process pressure maintained at 5 mTorr through a throttled gate valve. $PH_3$ is highly toxic and pyrophoric, and residual phosphorus deposits in the growth chamber can spontaneously ignite. Therefore, an $O_2$ purging step was implemented after each growth to remove any unreacted phosphorus from the growth platen before transferring the sample to the loadlock. As a result, traces of O incorporation in the films may be anticipated. The safety requirements and precautionary procedure for working with a $PH_3$ chamber are explained in a previous article [22] and in SI Section 1. Compositionally uniform films were achieved by rotating the platen during growth. Crystalline $CaZn_2P_2$ and $SrZn_2P_2$ thin films were grown on an a-$SiO_2$ (fused silica) at 200 °C with a deposition time of 2 h yielding the thickness of ~ 500 nm. The crystalline nature of the films was confirmed via high resolution synchrotron grazing incidence wide angle X-ray scattering (GIWAXS). Measurements were performed at beamline 11-3 at the Stanford Synchrotron Radiation Lightsource. A Rayonix 225 area detector was used to collect data at room temperature using an X-ray wavelength of $\lambda$ = 0.97625Å and an incident angle of 3°. See Fig. S5 for PXRD patterns of the compounds investigated here.

**Photoluminescence (PL) analysis**

Synthesized powdered samples of $BaCd_2P_2$, $CaCd_2P_2$, $SrCd_2P_2$, and $BaCd_2As_2$, were compressed into pellets for PL analysis. Additionally, PL data were collected from $SrZn_2P_2$ and $CaZn_2P_2$ thin films, ≈500 nm thick, grown on a-$SiO_2$ substrates. PL spectra were recorded using a Horiba Labram HR Evolution Raman spectrometer equipped with a micro-PL setup. The measurements employed a 50× objective lens and a 300 gr/mm grating, conducted at a temperature of 298 K to probe the band to band transitions for these materials. A 532 nm (~2.33 eV) laser source with a power output of 100 mW, attenuated by a 1% neutral-density (ND) filter, was used. Considering the optical efficiency of the equipment was determined to be ~30%, the effective laser power at the sample was approximately 0.3 mW. PL spectra were collected within the spectral range of 650-1050 nm (1.91-1.18 eV) for the $BaCd_2P_2$, $CaCd_2P_2$, and $SrCd_2P_2$ samples. For the $BaCd_2As_2$ sample, the spectral range was extended to 1100-1800 nm (1.12-0.69 eV) to accommodate its lower energy band transitions using InGaAs charge-couple device (CCD) detectors. For the $SrZn_2P_2$ and $CaZn_2P_2$ thin films, measurements were performed in the same spectral range as the powdered samples to facilitate comparison. To ensure the reliability of the data and mitigate site-specific variations, multiple regions on each sample were measured.

**Results and Discussion**

Despite the growing interest of the $AM_2Pn_2$ Zintl compounds, not all elemental combinations have been systematically explored. Isoelectronic mutation with A= Ca, Sr, Ba, Yb, Mg; M= Mn, Zn, Cd, Mg; and Pn=N, P, As, Sb, Bi amounts to 100 total possible compositions. We note that some Eu-containing $AM_2Pn_2$ compounds have been experimentally reported, but we have chosen to exclude these from our investigation due to convergence issues with Eu [37–41]. Querying the Materials Project for $AM_2Pn_2$ returns 89 $AM_2Pn_2$ entries in five unique crystal structure prototypes, with space groups $P\bar{3}m1$, *I4/mmm, Pnma, I4/mcm,* and *C2/m.* Upon further inspection the *I4/mcm,* and *C2/m* are entirely hypothetical for the $AM_2Pn_2$ with no experiments confirming their existence, so they will be excluded from further consideration. Compounds with the same stoichiometry as $AM_2Pn_2$ have been confirmed in the $R\bar{3}m$ and $P4/mmm$ structure type (such as $NaSn_2As_2$[42] and $BaPt_2P_2$[43], respectively), however they all contain elements beyond those considered here and do not appear within the $AM_2Pn_2$, so we exclude them as well. The $P\bar{3}m1$, *I4/mmm,* and *Pnma* structure prototypes are shown in Figure 1. Of those 89, 63 of the $AM_2Pn_2$ reported on the Materials Project are in the $P\bar{3}m1$ space group. Here, we report on our stability analysis and band structure for all of these potential compositions.

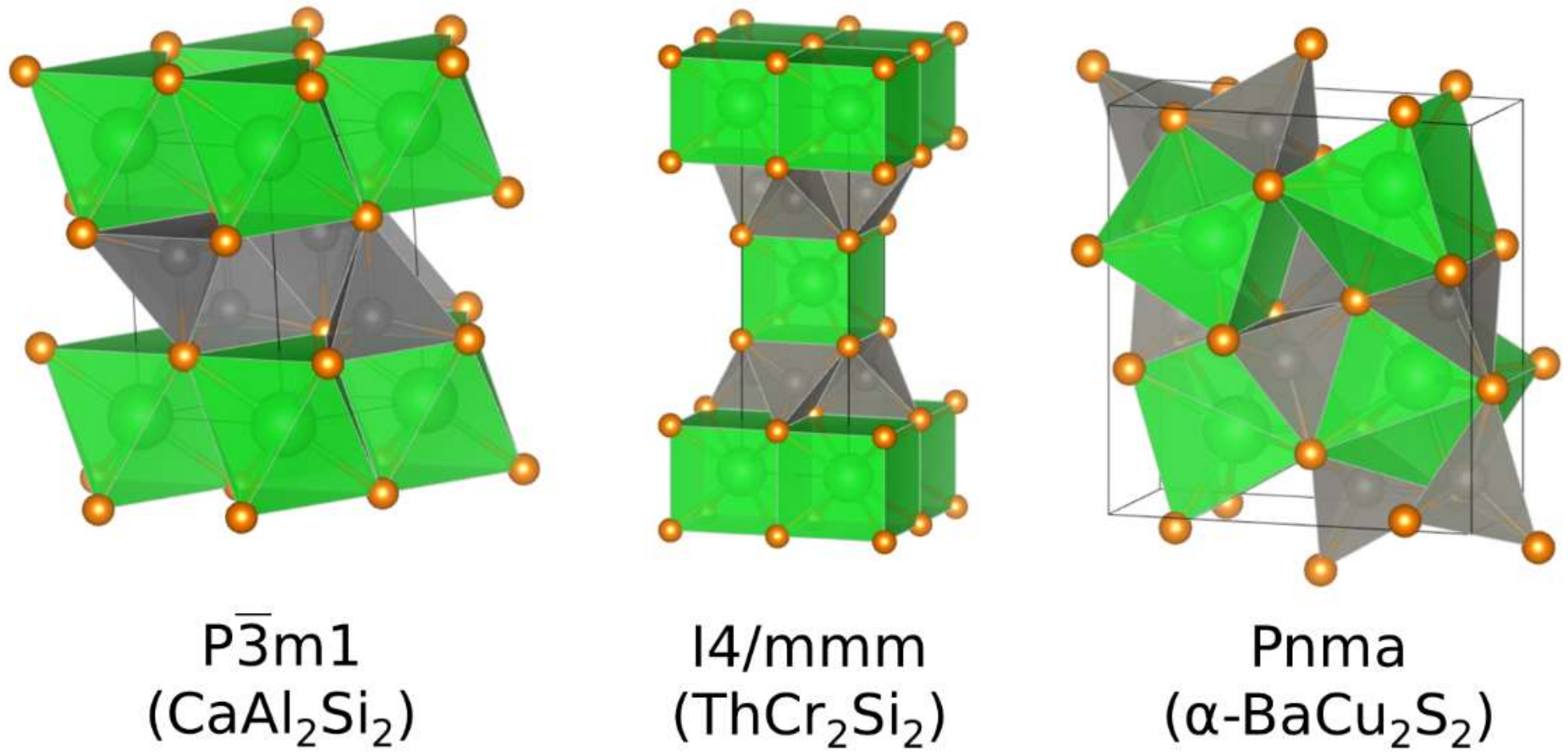

Fig. 1- The three crystal structure prototypes for the $AM_2Pn_2$ compounds used in the ground state structure search. Drawings produced with VESTA[44].

Fig. 2 shows the thermodynamic stability, described by the hull energy, of the $AM_2Pn_2$ compounds. The color scale is $E_{hull}$ as shown in Eq. (2); negative values represent the $AM_2Pn_2$ is stable, whereas positive values represent thermodynamically favorable decomposition into competing phases. The data points shaped as circles indicate that the ground-state structure for the $AM_2Pn_2$ is found to be the $P\bar{3}m1$ structure, while the square data points correspond to a ground state in one of the other structure types found from the Materials Project shown in Figure 1. See Fig. S3 for full details of the relative stability of other types. From Fig.1, it is visible that a large portion of possible compositions are predicted to be stable for isoelectronic substitution across the A and M sites. In addition, most of the $AM_2Pn_2$ do favor the $P\bar{3}m1$ space group. The exceptions to these are $AM_2N_2$ where only a few $AMg_2N_2$ and $AZn_2N_2$ are stable. The compounds with the cation A-site occupied by Mg are mostly unstable, despite being stable in well-characterized thermoelectric materials such as $Mg_3Sb_2$ (or $MgMg_2Sb_2$) and $Mg_3Bi_2$ ($MgMg_2Bi_2$). In our screening we additionally evaluate the binary $Mg_3Pn_2$ in the $Ia\bar{3}$ space group structure since this was also reported on the Materials Project, but does not have a ternary analog. We note that $Mg_3Sb_2$ has been identified to be close to a dynamical instability due to the size mismatch between Mg and Sb in the $P\bar{3}m1$ space group [45]. The ground state structures of $BaZn_2Pn_2$ and $BaMn_2Pn_2$ are not $P\bar{3}m1$ which is understood from the large size difference between the A and M sites in these two compounds[46].

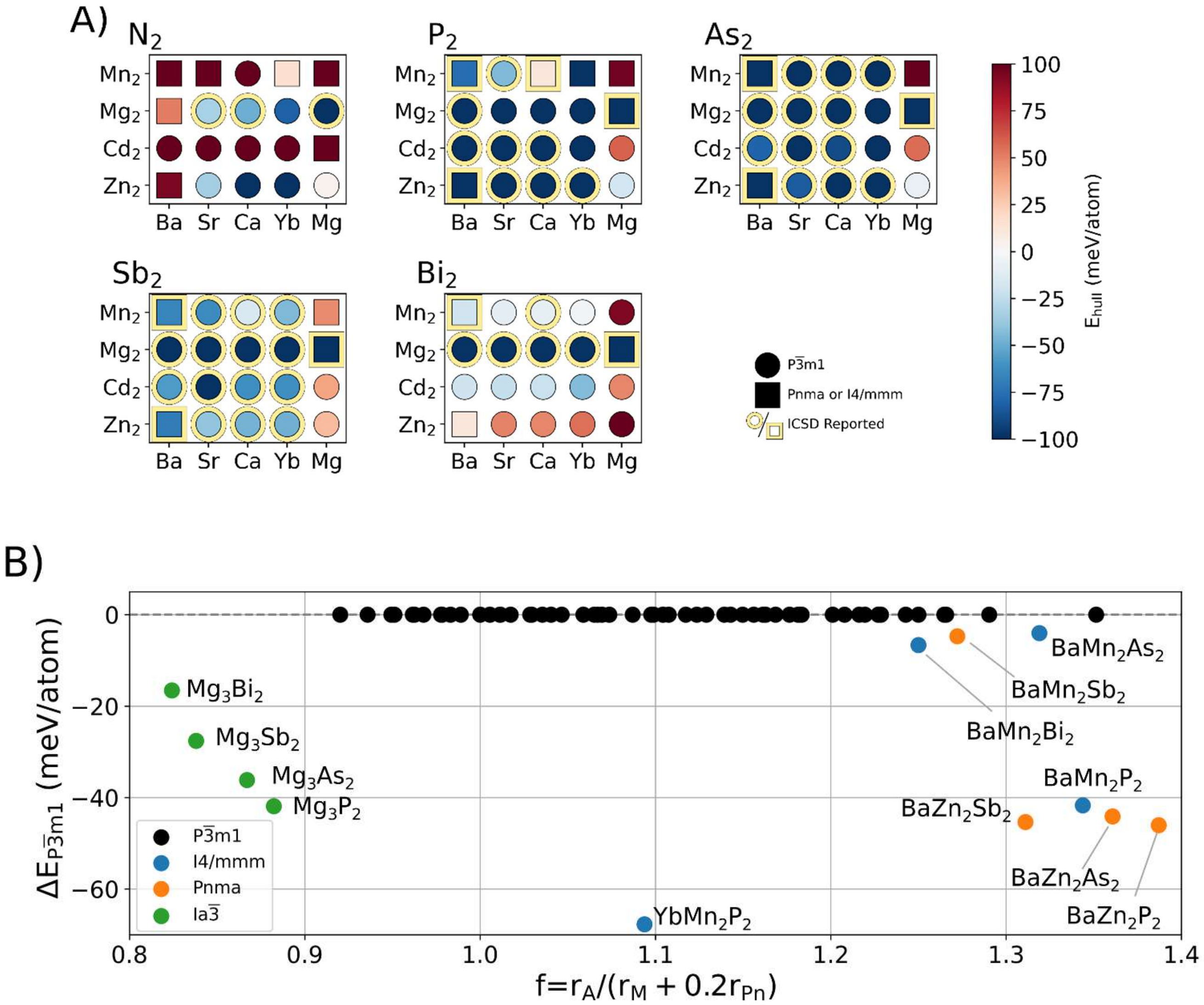

Fig. 2 - A) Stability of the $AM_2Pn_2$. Matrix showing the $r^2$SCAN predicted energies for the formation of each $AM_2Pn_2$ from its decomposition products. Here, a negative $E_{hul}$ (blue) indicates stability, and positive (red) indicates instability. Circular data points represent compositions where $P\bar{3}m1$ is the ground state structure, and squares represent cases where the ground state is one of the other two structures presented in Fig. 1. Compositions reported in the ICSD 5.3.0 (ICSD release 20241001-1145) are bordered in yellow. See Table S1 for tabulated data from this plot. B) The empirical size correlation of Klufers et al. [47] versus calculated energy difference between the most stable polymorph and the $P\bar{3}m1$ polymorph, $\Delta E_{P\bar{3}m1}$, for stable compounds only ($E_{hull} < 0$). A plot of all compounds in this investigation is in Fig. S2. Color of the point represents the ground state space group of the composition. For clarity, text labels of the $P\bar{3}m1$ phases have been removed.

In Fig. 2b we compare our calculated data to the empirical stability rule based on atomic radii proposed by Klufers et. al [47]:

$$f = \frac{r_A}{r_M + 0.2 r_{Pn}} \quad (5)$$

We find that our data matches the trend of Peng et. Al [46] in that there is an approximate threshold $f > 1.25$ above which the $P\bar{3}m1$ structure type is less likely to be stable compared to *I4/mmm* or *Pnma* structure types. We additionally included the $Mg_3Pn_2$ compositions which allows access to lower $f$ values and shows that there is a lower threshold for $f < 0.9$ where the $Ia\bar{3}$ structure type is stabilized over $P\bar{3}m1$ at 0K. In Peng et al.[45] it is noted that $P\bar{3}m1$ $Mg_3Sb_2$ is close to a dynamical instability which gives this phase a high vibrational entropy. Coupled with the small difference in energy for the $P\bar{3}m1$ and $Ia\bar{3}$ phases of $Mg_3Pn_2$ as shown in Fig. 2b, these phases form $P\bar{3}m1$ at high temperatures. The proximity of the $Mg_3Sb_2$ to the threshold further rationalizes these results. This is a relatively wide window of $f$ for which the $P\bar{3}m1$ is favored, explaining why the majority of phases favor formation in this structure. The implications of Eq. (5) are that the pnictogen ion has a relatively small impact on the structural stability of the $AM_2Pn_2$. For a combination of the largest A ion, Ba, and relatively small Zn and Mn, $f$ increases above the threshold, leading to the stabilization of other structure types.

Among the 100 compositions investigated, 54 are recorded in the ICSD[48]. Fifty out of these 54 experimentally observed compounds, match the ICSD records in terms of the existence of a compound of that composition (showing its stability) and the compound's reported space group, demonstrating the predictive power of our approach. $CaMn_2P_2$ has a discrepancy, for which we compute a slightly positive $E_{hull}$ of 12 meV/atom. This compound has been reported experimentally several times crystallizing in the $P\bar{3}m1$ space group structure[6,49–52]. The discrepancy could come from vibrational or magnetic configuration effects. $Mg_3Sb_2$, $Mg_3Bi_2$, and $Mg_3N_2$ are all correctly predicted to be stable, but not in the correct space groups. $Mg_3Sb_2$ and $Mg_3Bi_2$ are predicted to crystallize in the $Ia\bar{3}$ space group structure, whereas they are reported in the $P\bar{3}m1$ space group structure. $Mg_3Sb_2$ in the $P\bar{3}m1$ space group structure has been noted to be near a dynamical instability, giving it a large vibrational entropy[45]. At high synthesis temperatures, the entropy could overcome the small energy difference of 28 meV/atom between the $Ia\bar{3}$ and $P\bar{3}m1$ space group structures, as seen in Fig. S3. $Mg_3Bi_2$ could be expected to have a large vibrational entropy as well since it also has Mg on the A site like $Mg_3Sb_2$, which has been attributed as the cause of the lattice instabilities. $Mg_3Bi_2$ has an even smaller energy difference of 17 meV/atom between the $Ia\bar{3}$ and $P\bar{3}m1$ space group structures. The calculation for $Mg_3N_2$ in the $Ia\bar{3}$ space group structure did not converge so its stability could not be predicted.

There are a number of compounds we predict to be stable that are not found in the ICSD. In the literature, some of these have actually already been synthesized experimentally with reported PXRD patterns, including $SrZn_2N_2$[53], $CaZn_2N_2$[54,55], $MgZn_2P_2$[56], $YbCd_2As_2$[57] and $SrMn_2As_2$[58]. The remaining compounds are new compounds not synthesized yet but should be synthesizable ($E_{hull} < 0$) according to our calculations: $YbMg_2N_2$, $YbZn_2N_2$, $CaMg_2P_2$, $SrMg_2P_2$, $YbMg_2P_2$, $YbCd_2P_2$, $YbMn_2P_2$, $YbMg_2As_2$, $MgZn_2As_2$, $BaCd_2Bi_2$, $CaCd_2Bi_2$, $SrCd_2Bi_2$, $YbCd_2Bi_2$, $SrMn_2Bi_2$, and $YbMn_2Bi_2$. To our knowledge, $SrMg_2P_2$[59], $CaMg_2P_2$[60], $BaCd_2Bi_2$[61], $CaCd_2Bi_2$[62], $SrCd_2Bi_2$[63], and $SrMn_2Bi_2$[64] have previously been studied computationally.

For the rest of our discussion focusing on the electronic structure we will only consider the $AM_2Pn_2$ composition in the $P\bar{3}m1$ structure type because the majority of compounds are stable in this structure type. [58,65–70]We will also exclude the manganese compounds ($AMn_2Pn_2$) from further discussion as their computed bandgap can vary by more than 1 eV depending on their magnetic configuration, while their bandgaps have been measured to be ~10 meV which will be difficult to capture given the typical precision of first-principles methods[58,65–70]. Fig. 3 shows the bandgaps for the $P\bar{3}m1$ $AM_2Pn_2$ compounds. Relatively few of them have experimentally reported bandgaps. We note that we computed the bandgaps using HSE+SOC based on the PBE-relaxed structures. Fig. 3 shows that the $P\bar{3}m1$ $AM_2Pn_2$ materials have a wide range of bandgaps, spanning from 0 to over 3 eV. A table of the bandgaps for $AM_2Pn_2$ compounds is available in Supplemental Table S2 and full band structures can be found in Fig. S4. As commonly observed, the bandgaps decrease moving down the group of pnictides from nitrides to bismuthides. For a given anionic metal, Mg on the M-site has the largest bandgaps. Zn and Cd containing analogs often have similar bandgaps, but Cd more often has a direct bandgap and Zn is usually indirect. The bandgap is not strongly affected by the element on the A site. The wide range of bandgaps means a wide range of potential applications. The $AM_2Pn_2$ materials are already recognized as promising thermoelectric materials, but this range of bandgaps and the recent results on the defect tolerance of $BaCd_2P_2$ and $CaZn_2P_2$ suggest they can also make promising infrared detectors, single-junction solar cells, tandem solar top cells, light emitting diodes (LEDs) or photoelectrocatalysts [21,22]. In a later section we will discuss how candidate compounds can be used in specific applications.

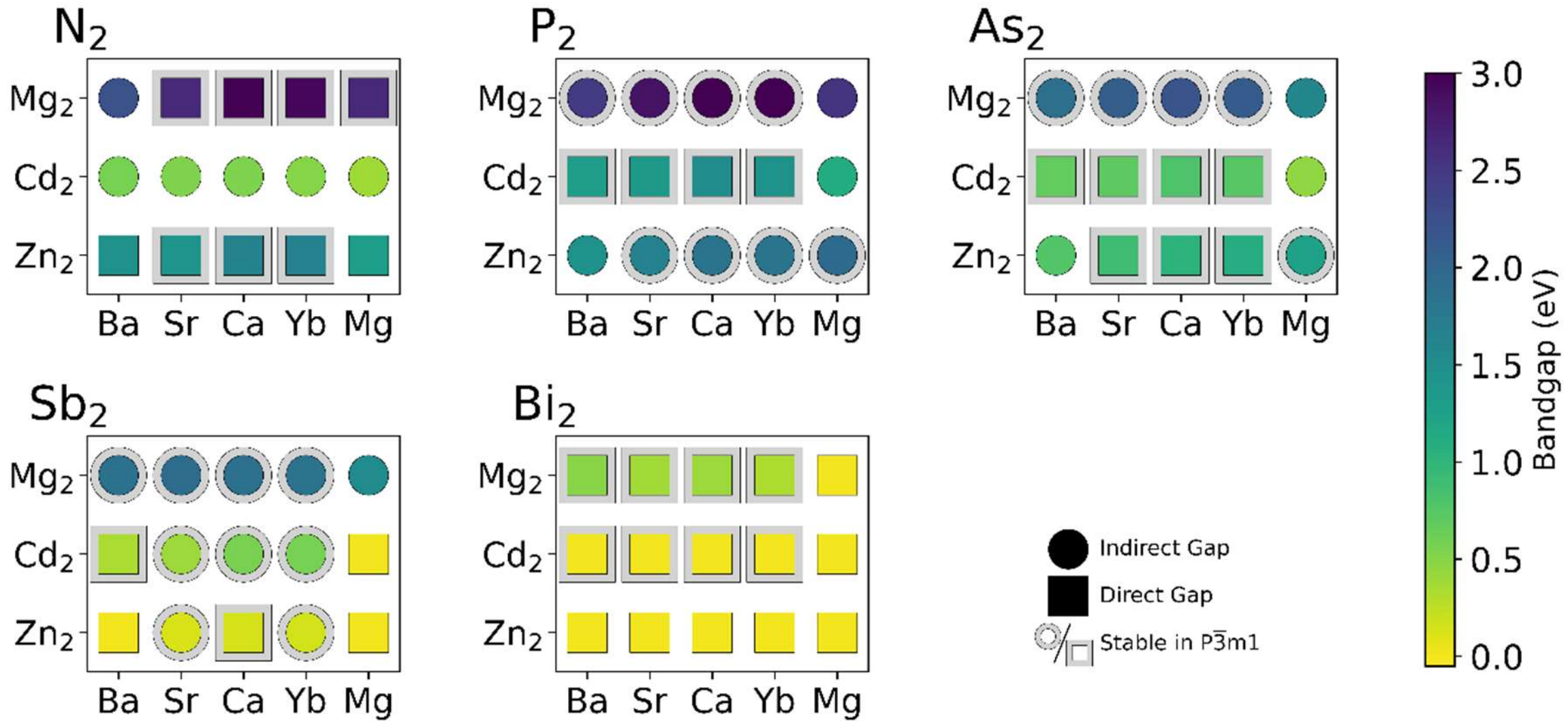

Fig. 3 - HSE+SOC bandgaps of $AM_2Pn_2$ in the $P\bar{3}m1$ structure. The data point shape denotes the nature of the bandgap, squares are direct bandgap materials and circles are indirect. See Fig. S4

for full band structures. Compositions stable ($E_{hull} < 0$) in the $P\bar{3}m1$ spacegroup structure are bordered in gray.

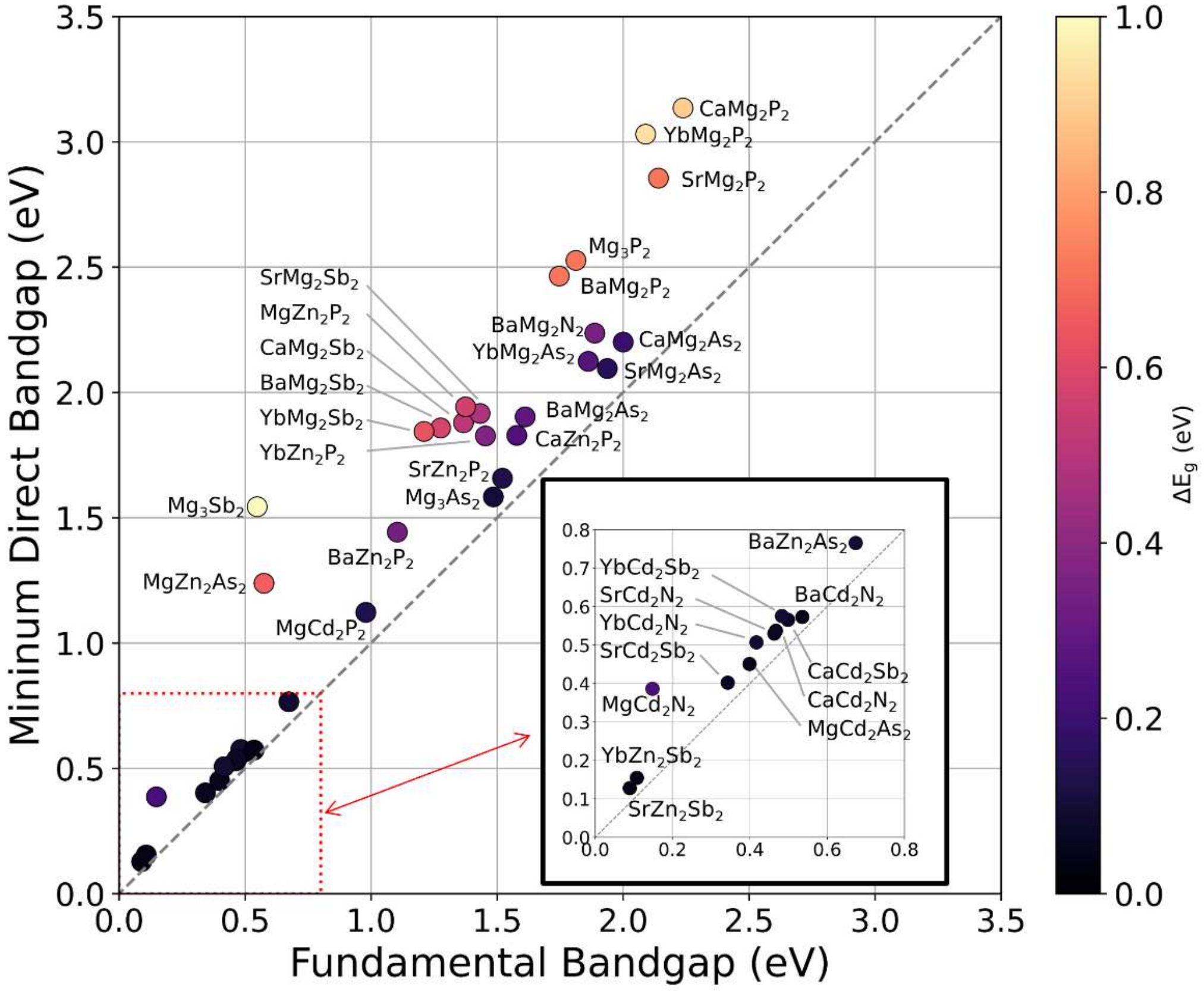

Fig. 4 - Fundamental vs minimum bandgap of the $P\bar{3}m1$ $AM_2Pn_2$ materials. Direct bandgap materials are omitted because they have $\Delta E_g = 0$ by definition. The color scale is the difference between the direct and indirect bandgap, $\Delta E_g$. Small bandgap materials are shown as inset for clarity. The x=y line (gray, dashed) is shown as a visual reference.

Rather than simply categorizing each material as direct or indirect, we have quantified the difference between these bandgaps, $\Delta E_g$, shown in Fig. 4. When considering potential applications for materials, especially in optoelectronics, those with an indirect bandgap are not optimal, but a small $\Delta E_g$ can be tolerable[71]. For example, in a thin film solar absorber the optical absorption is primarily due to direct band-to-band transitions, whereas the open circuit voltage, $V_{oc}$ is at most the minimum bandgap, so a $\Delta E_g > 0$ represents a decrease in optical absorption for a given $V_{oc}$. Twelve materials in this study have $0 < \Delta E_g < 0.1\ eV$, meaning they are somewhat close to a direct material. Of these nearly direct materials, $Mg_3As_2$ has the largest indirect bandgap (1.49 eV) while all other nearly direct materials have bandgaps less than 0.7 eV. $Mg_3Sb_2$ is the most indirect material, with a 0.55 eV indirect bandgap and a 1.54 eV direct bandgap. The subfamily $SrMg_2P_2$, $YbMg_2P_2$, and $CaMg_2P_2$ are also highly indirect with $\Delta E_g = 0.72$, 0.93, and 0.89 eV, respectively.

From our band structure calculations, we have also extracted the estimated average carrier conductivity effective masses for the $P\bar{3}m1$ $AM_2Pn_2$ materials. Effective masses set carrier mobility which is important in PV or thermoelectric applications, where low effective masses are desirable for both applications. From Fig. 5a it is seen that the class of materials studied here has relatively low electron masses, typically below about 1.0 $m^*_e$, and about 0.5 $m^*_e$ on average. Some exceptions to this are $BaZn_2As_2$, $SrZn_2As_2$, $BaCd_2Sb_2$, and $SrZn_2Sb_2$ which have much higher electron effective masses. On the other end of the spectrum, the nitrides have quite low effective masses, less than 0.25 $m^*_e$ on average. For the cation M site, Cd generally leads to lower electron effective masses than Mg or Zn counterparts. For holes there is more variation in the effective mass throughout the family. Fig. 5b shows nitrides have the heaviest holes. Other pnictides have a fairly consistent hole effective mass of ~0.75 $m^*_e$. We note that for thermoelectric applications it is the majority carrier transport that is most important, while photovoltaics, which involve optical excitations, tend to depend on minority carrier transport. For example, Si solar cells use p-type substrates such that the minority carriers have relatively high mobility.

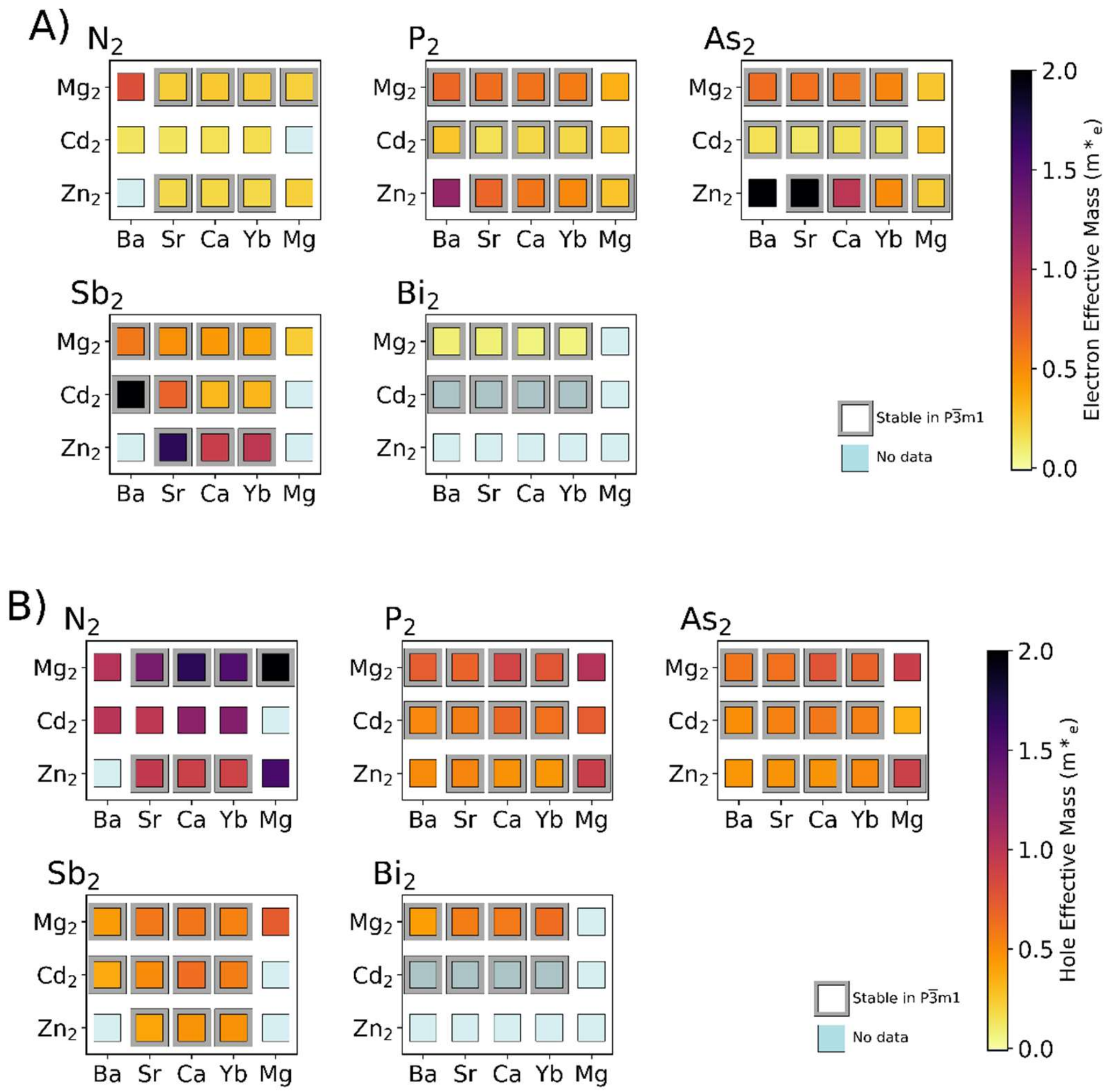

Fig. 5 – Average conductivity effective masses of A) electrons and B) holes for the $P\overline{3}m1$ $AM_2Pn_2$. Materials with computed bandgaps less than 0.1 eV have been excluded. Light blue represents no data, either because of exclusion or no completed calculations. Compositions stable ($E_{hull} < 0$) in the $P\overline{3}m1$ spacegroup are bordered in gray.

Here, we compare our calculated bandgaps to experimental data available. We note that the HSE+SOC bandgaps are based on PBE-relaxed structures. Overall, we find excellent agreement between the calculated bandgaps and the experimental reports as shown in Fig. 6. Excluding specific measurements for $CaZn_2P_2$ [72], $SrCd_2As_2$ [73], and $BaCd_2Sb_2$ [74] (discussed below), the mean average error (MAE) is 0.16 eV. This is consistent with other reports on the agreement between full HSE calculations and experimental bandgaps [75,76]. This gives us confidence in our results and that they can be used predictively to apply to the materials not yet well characterized. The most overestimated bandgap is $CaMg_2Bi_2$, which we calculate is 0.40 eV direct bandgap but is reported to be 0.2 eV by electrical resistivity measurements by May et al[77], although this measurement is likely a lower limit for the bandgap since the technique used may be measuring the transition between the valence band and a mid-gap defect state if present. Conversely the most underestimated bandgap is $SrCd_2Sb_2$ which we predict should have an indirect bandgap of 0.34 eV and a minimum direct bandgap of 0.40 eV, but it is measured as 0.63 eV by Jin et al[78]. with optical measurements on a Tauc plot. So, in general, our results match experiments well, and should be applicable predictions for those systems that have not yet been investigated and reported experimentally.

As mentioned previously, we find several reports that neither match our calculations, nor other published bandgap values. We summarize those findings here and present a full discussion in SI Section 3. In the case of $CaZn_2P_2$ there have already been several independent measurements of its bandgap, so by direct comparison we can eliminate reports by Ponnambalam et al. of a 0.6 eV bandgap[72]. It is still uncommon to have several bandgap measurements of the same material in this class of compounds, but by looking at the trends in bandgap in groups of similar compounds we notice several inconsistencies. For example, Chen et al. [73] measured a 0.21 eV bandgap for $SrCd_2As_2$ which is less than reports of the bandgaps for $BaCd_2As_2$ and $SrCd_2Sb_2$, though it is expected to be higher than either of these from our calculations as well as their trend in atomic mass. It appears our calculations underestimate the bandgaps of $CaMg_2N_2$ and $CaZn_2N_2$, likely due to overestimation of the lattice parameter from the initial PBE relaxation. Fully HSE calculations can correct this error, but at an increased computational cost.

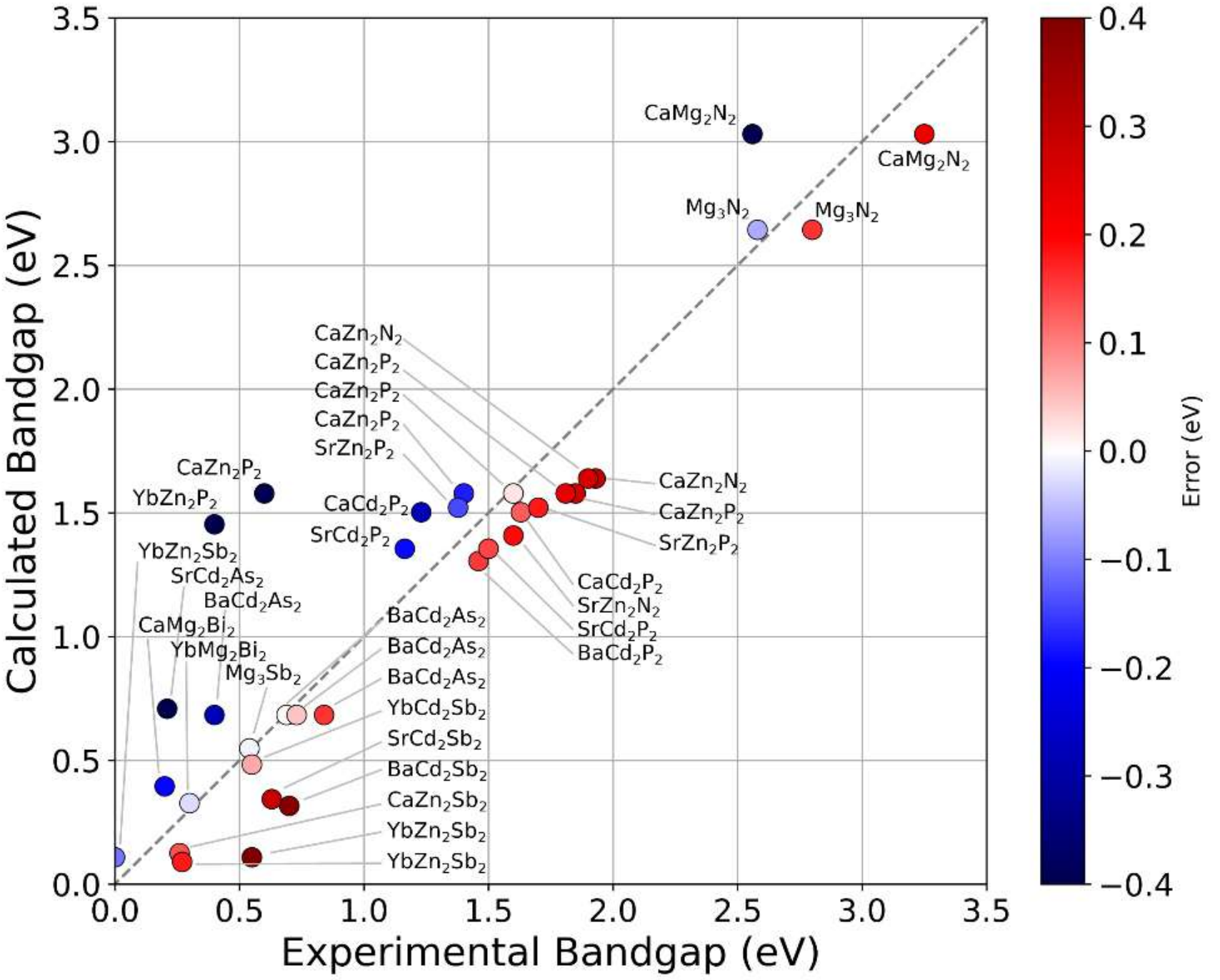

Fig. 6 - Comparison of experimentally reported bandgaps and our calculations. The color of the point represents the difference between the two values. The x=y line (gray, dashed) is shown as a visual reference.

*Experimental Verification*

We have synthesized $BaCd_2P_2$, $CaCd_2P_2$, $SrCd_2P_2$, and $BaCd_2As_2$ powders and $CaZn_2P_2$ and $SrZn_2P_2$ thin films and performed photoluminescence spectroscopy studies, shown in Fig. 7, to obtain their bandgaps. The measured (calculated) direct bandgaps for $BaCd_2P_2$, $CaCd_2P_2$, $SrCd_2P_2$, $BaCd_2As_2$, $CaZn_2P_2$ and $SrZn_2P_2$ are 1.46 eV (1.31 eV), 1.58 eV (1.50 eV), 1.29 eV (1.35 eV), 0.73 eV (0.68 eV), 1.93 eV (1.83 eV), and 1.81 eV (1.66 eV), respectively. For $BaCd_2P_2$, $CaCd_2P_2$, $SrCd_2P_2$, and $BaCd_2As_2$ we have assigned the highest peak to the band-to-band transition though $BaCd_2P_2$ does show a low intensity sub-bandgap defect peak at about 1.25 eV. For $CaZn_2P_2$, we have assigned the highest peak to the band-to-band transition corresponding to the direct bandgap. We attribute the second smaller peak centered at 1.65 eV to $CaZn_2P_2$'s indirect bandgap which we calculate to be 1.58 eV. These experimentally measured bandgap values agree well with our computational results, despite the slight systematic underestimation due to the inaccuracy of the lattice parameters expected from the PBE relaxation. To the best of our knowledge, our results are the first report of the experimentally measured bandgap for $CaCd_2P_2$ and $SrCd_2P_2$.

There are conflicting experimental reports of the bandgap of $BaCd_2As_2$. In Kunioka et al.[79] the bandgap is reported to be 0.40 eV while, Yang et al.[80] reports their measurement of the bandgap is more than double that at 0.84 eV. Both measurements similarly estimate the bandgap from the change in resistivity with respect to temperature. Our calculation of 0.68 eV more closely matches Yang et al. [80], but is close to the middle of the two measurements, so makes it difficult to discern which report is more accurate. Now, with our optical data we report that our experimental measurement of the bandgap is 0.73 eV which matches very closely with our calculation and agrees better with the bandgap reported in Yang et al.[80]. This clarifies the bandgap of $BaCd_2As_2$ with greater accuracy.

To verify our prediction of new (i.e. previously unreported) compounds, we have tried to synthesize $SrCd_2Bi_2$. From Fig. 2, the $ACd_2Bi_2$ compounds are predicted to be stable with an $E_{hull}$ that is only slightly negative, yet they have not been reported. Unfortunately, our solid-state reactions do not result in the target compound, but instead $SrCdBi_2$ + Cd. This can be rationalized given the small $E_{hull}$ (-24 meV/atom) for $SrCd_2Bi_2$ which has a predicted decomposition of $SrCdBi_2$ + Cd. In chemical synthesis, the formation of target compounds with small decomposition energies can be kinetically controlled to form other metastable products [81]. We have not tested alternate synthetic routes for $BaCd_2Bi_2$, such as starting from binary precursors rather than elemental precursors, which may still allow the formation of this new compound.

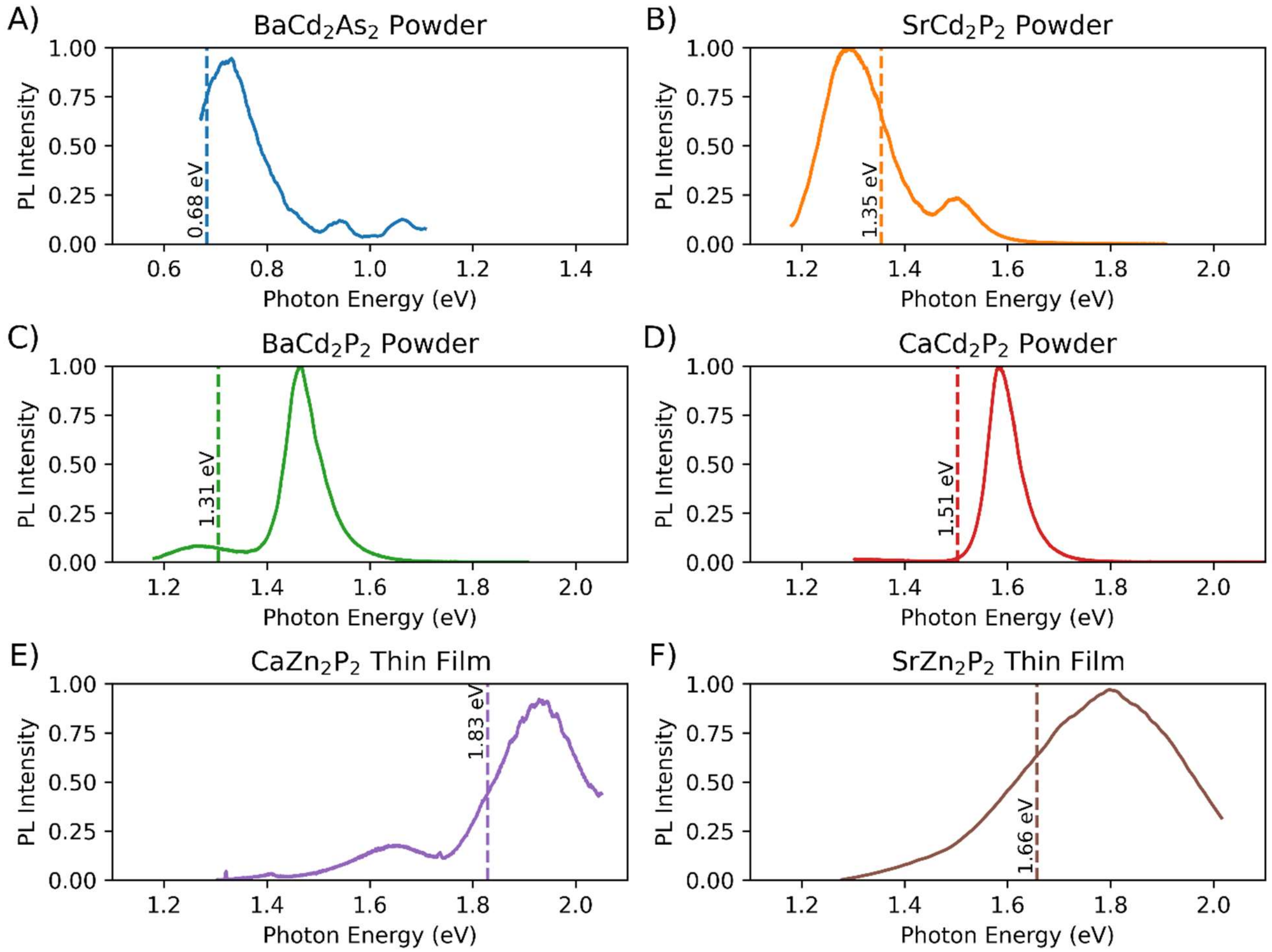

Fig. 7 - PL Spectra of select $AM_2Pn_2$. A-D) powder samples of $BaCd_2As_2$, $SrCd_2P_2$, $BaCd_2P_2$, and $CaCd_2P_2$, respectively. E and F) Thin film samples of $CaZn_2P_2$ and $SrZn_2P_2$. Vertical dashed lines represent the HSE+SOC calculated direct bandgaps for visual reference to compare between experiment and computations. PL spectrum of $CaZn_2P_2$ is taken from Quadir et al. [22]

*Candidate Selection for Specific Applications*

The electronic structure of the $AM_2Pn_2$ can be widely tuned by varying chemistry. We now discuss what chemistries could work best for certain applications, including thermoelectric devices, infrared detector, and photovoltaics.

The most widespread application considered for $AM_2Pn_2$ compounds is thermoelectric materials. Thermoelectric applications favor small bandgap semiconductors as it correlates with potential for high doping and electrical conductivity. It also requires low lattice thermal conductivity which is easier to reach with heavier elements. In that respect, the antimonides present the most adequate chemistry and they have indeed been widely studied in the thermoelectric field with $EuZn_2Sb_2$ [82], $YbCd_{2-x}Zn_xSb_2$ [83], and most notably with $Mg_3Sb_2$ [19,20,84,85]. Another desirable property for thermoelectric materials is low effective masses combined with valley degeneracy. Valley degeneracy is enhanced by conduction or valence band pockets off the $\Gamma$ point due to their higher multiplicity in the Brillouin zone[86]. The valence bands are centered in $\Gamma$ for all of the antimonides, indicating that the electronic structure of a p-type thermoelectric in the $AM_2Pn_2$ will not favor a high zT even if doping is achievable. On the other hand, the conduction band can have off-$\Gamma$ pockets. $Mg_3Sb_2$ clearly has the highest valley degeneracy of all $AM_2Pn_2$ compounds with conduction band pockets at the *M*, *K*, and *L* points (Fig. S4). Other $AMg_2Sb_2$ also have conduction band valley degeneracy, though to a lesser extent than $Mg_3Sb_2$. The effect of Mg on the band structure is to raise the conduction band at the $\Gamma$ point, while keeping other local minima of the conduction band the same, as can be seen, for example, in the band structures of $CaCd_2Sb_2$ versus $CaMg_2Sb_2$ (see Fig. S4). This matches the findings of Zhang et al.[20] that, among the $AM_2Pn_2$, a multi-valley conduction band is unique to the binary $Mg_3Pn_2$ compounds. These considerations do not account for whether it is actually possible to properly dope the material, but just show that *n*-type doping is more promising than *p*-type doping.

Another emerging application for the $AM_2Pn_2$ is in PV. Notably, $BaCd_2P_2$ has shown a lot of potential as a single junction solar absorber material[21]. Defects, especially deep traps, can limit the performance of solar absorbers. Defect computations and carrier lifetime measurements have shown favorable defects in the $AM_2Pn_2$ family, at least for $BaCd_2P_2$ and $CaZn_2P_2$ [21,22]. Beyond single junction materials, solar absorbers for tandem cells are of great interest. If combined with silicon bottom cell of bandgap 1.1 eV, a top cell with a bandgap of ~1.5-1.8 eV [87] is most efficient. Our analysis shows a large range of phosphides, nitrides and arsenides with suitable bandgaps for single-junction or tandem solar cells. Many of the $AMg_2P_2$ or $AMg_2As_2$ have large bandgaps so may seem promising, but we recognize that these compounds may not be air- or moisture-stable due to the Mg content[88]. Fewer nitrides form the $AM_2Pn_2$ composition than other pnictogens, but the Zn nitrides could be an option because of their suitable bandgaps and a low electron effective mass. $CaZn_2N_2$ is especially appealing as highlighted by previous reports[54,55] which suggest its use as a solar absorber in single junction photovoltaic cell, though we find its bandgap is better suited for tandem top cells. We note that consideration about doping will be important to account for when studying materials for solar cells as well. Computed defects in phosphides indicate that it will be very difficult to produce *n*-type $BaCd_2P_2$ and $CaZn_2P_2$. In general, $AM_2P_2$ compounds are likely to be insulators or *p*-type semiconductors. $SrZn_2N_2$ and $CaZn_2N_2$ are both weakly *n*-type semiconductors [53,55].

The figure of merit for infrared detectors relates quite well to PV materials [89,90]. Long carrier lifetime and high optical absorption are similarly sought after but in lower bandgap materials. The favorable defects and absence of low formation energy, deep traps in phosphides could extend to smaller bandgap materials such as antimonides. Interesting candidates include $SrCd_2Sb_2$ and $CaZn_2Sb_2$, both of which are thermodynamically stable and show a small direct bandgap of 0.42 eV and 0.14 eV, respectively. Since the bismuthides form metallic systems with similar $P\bar{3}m1$ structures, it could be attractive to alloy some of the isostructural antimonides with bismuth to tune the bandgap to smaller values and make absorbers for long wavelength IR detection as performed in the $Hg_{1-x}Cd_xTe$ (MCT) system[91,92].

Finally, we note that given the broad stability of this class of materials and that most form in the same $P\bar{3}m1$ structure, alloys of these various ordered compounds may form quite readily. This could open the way to tune the bandgap and electronic structure to achieve the highest performance materials in thermoelectrics, photovoltaics, infrared detectors and other applications[85].

*Conclusions*

The $AM_2Pn_2$ class of materials has the remarkable ability to form stable compounds in a large number of compositions, for which they are only just starting to have their full utility realized. In this work we have computationally explored the family of the $AM_2Pn_2$ Zintl compounds, composed of 100 compositions. We clarified their thermodynamic stability and equilibrium structure, showing that most of them are stable in the $P\bar{3}m1$ space group. Fifteen new stable $AM_2Pn_2$ compounds are predicted. We have studied the electronic band structures of $P\bar{3}m1$ $AM_2Pn_2$ compounds, revealing that the bandgap of this class of materials spans a wide range from metals to semiconductors above 3 eV. Given the stability and diversity of this compositional family, there may be a variety of applications for which these materials can be used, beyond the thermoelectric materials for which they are more traditionally recognized. Based on the stability and electronic structure, we have selected preliminary candidates for tandem top cell solar absorbers, infrared detectors, and thermoelectric applications. We hope that this work serves as a roadmap to inspire more in-depth explorations of the individual $AM_2Pn_2$ compounds.

*Acknowledgments*

This work was supported by the U.S. Department of Energy, Office of Science, Basic Energy Sciences, Division of Materials Science and Engineering, Physical Behavior of Materials program under award number DE-SC0023509 to Dartmouth and was authored in part by the National Renewable Energy Laboratory, operated by Alliance for Sustainable Energy, LLC, for the U.S. Department of Energy (DOE) under contract no. DE-AC36-08GO28308. All computations, syntheses, and characterizations were supported by this award unless specifically stated otherwise. This research used resources of the National Energy Research Scientific Computing Center (NERSC), a DOE Office of Science User Facility supported by the Office of Science of the U.S. Department of Energy under contract no. DE-AC02-05CH11231 using NERSC award BES-ERCAP0023830. A.P. acknowledges support from a Department of Education GAANN fellowship. The analysis on infrared detector materials was supported by the United States Air Force Office of Scientific Research under Award No. FA9550-22-1-0355. The views expressed in the article do not necessarily represent the views of the DOE or the U.S. Government.

*SI Section 1: Safety Warning*

*Warning: The starting reagent, alkaline-earth metals are air- and water-reactive and should be handled carefully in an inert atmosphere. At >400 °C inside the reaction ampoule, excessive vapor pressure of As or P as well as reaction of alkaline-earth metal with silica may compromise the silica ampoule resulting in shattering or explosion. The annealing steps must be conducted in a well-ventilated environment, such as in a fume hood.*

*Warning: We strongly emphasize that $PH_3$ is toxic and pyrophoric and P deposits leftover in a growth chamber can spontaneously combust during venting and routine chamber service (part changes, cleaning, etc.). Thus, additional safety controls, including robust interlocking, hydride gas monitoring, pump/purge cycling, self-contained breathing apparatus, flame-retardant personal-protective equipment, exhaust abatement, and others, must be rigorously implemented from the onset for any growth chamber intended to utilize $PH_3$ or prepare phosphide samples.*

*SI Section 2: Comparison of Stability calculated with PBE and $r^2$SCAN*

Preliminary tests showed that PBE systematically failed to show the stability of a series of compounds that have already been experimentally synthesized. Specifically, this happens for the series of compounds $AMn_2Pn_2$ where Pn = Sb, Bi. In $r^2$SCAN, the energy of these phases is dramatically decreased, showing them as stable, in agreement with the reports of the synthesis of these phases. Additionally, the $ACd_2Bi_2$ are stabilized, moving them further below the hull whereas they were previously very nearly on the hull.

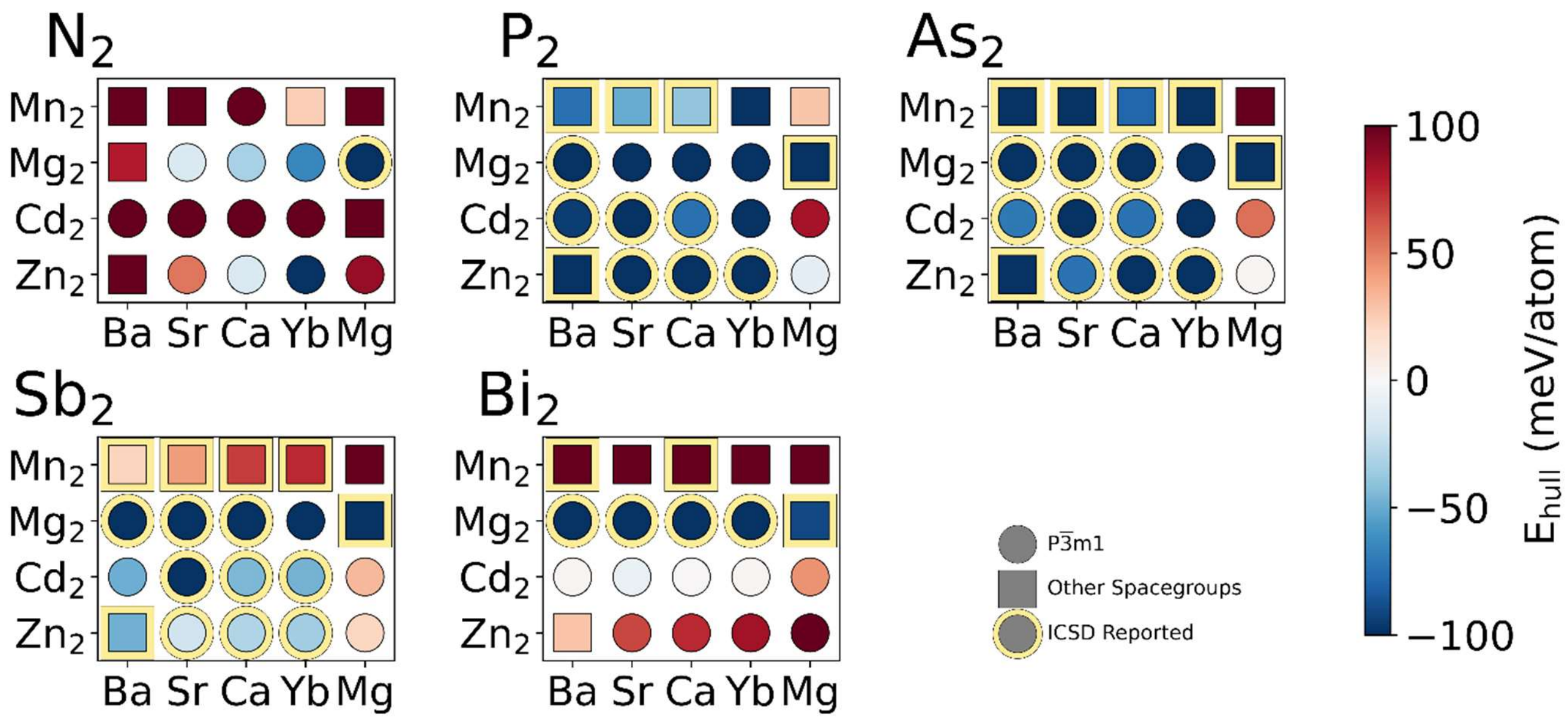

Fig. S1 - Stability of the $AM_2Pn_2$. Matrix showing the PBE predicted energies for each $AM_2Pn_2$ to decompose into its products. Here, a positive energy of decomposition (blue) indicates stability, and negative (red) indicates instability. Circular data points represent compositions where $P\bar{3}m1$ is the ground state structure, and squares represent where it is one other structure of the other four structures explored. Compositions reported on the ICSD are bordered in green.

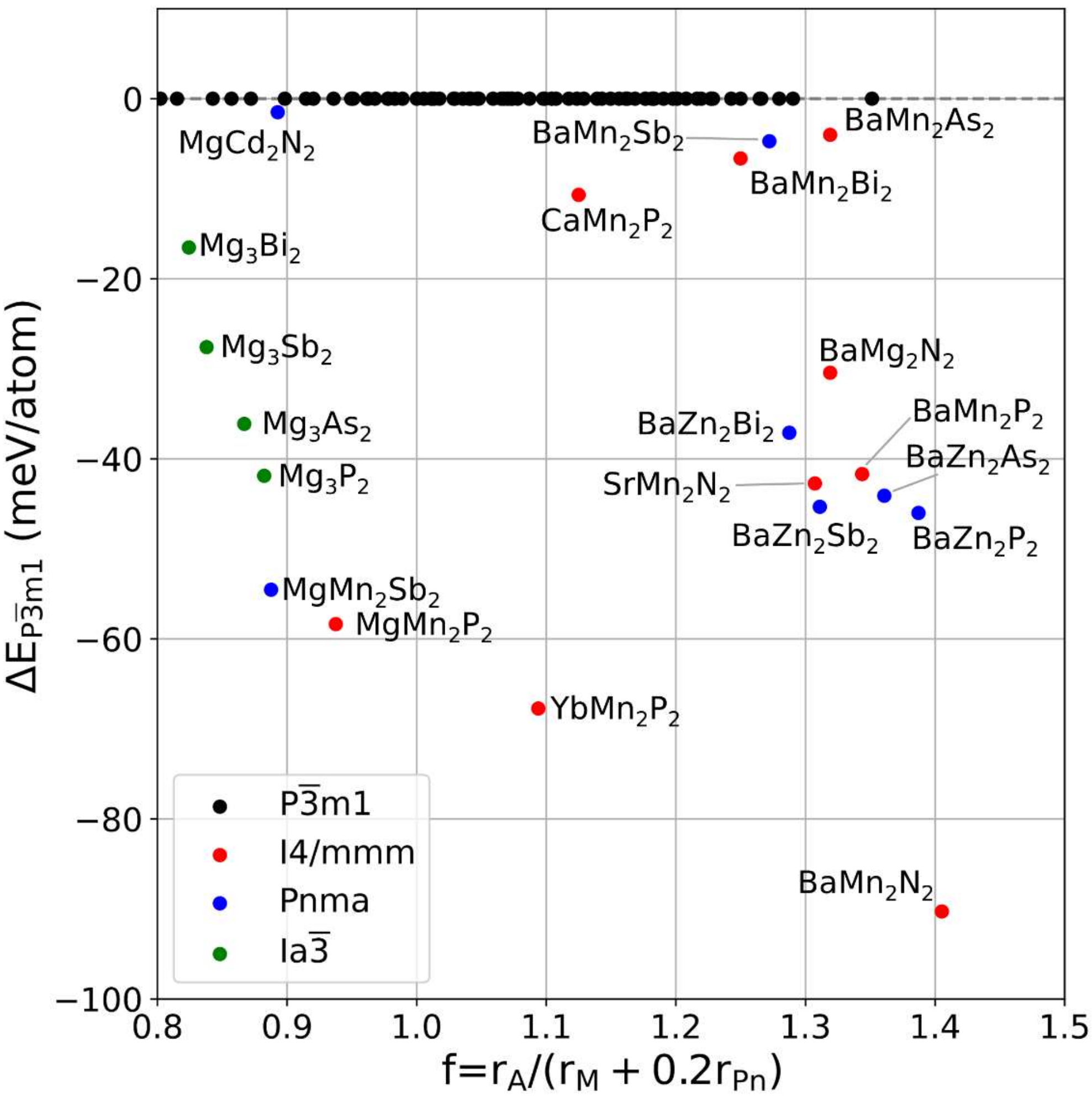

Fig. S2 - The empirical size correlation of Klufers et Al. [1] versus calculated energy difference between the most stable polymorph and the P-3m1 polymorph, $\Delta E_{P\bar{3}m1}$, for all compositions studied in this investigation. Color of the point represents the ground state space group of the composition. For clarity, text labels of the $P\bar{3}m1$ phases have been removed

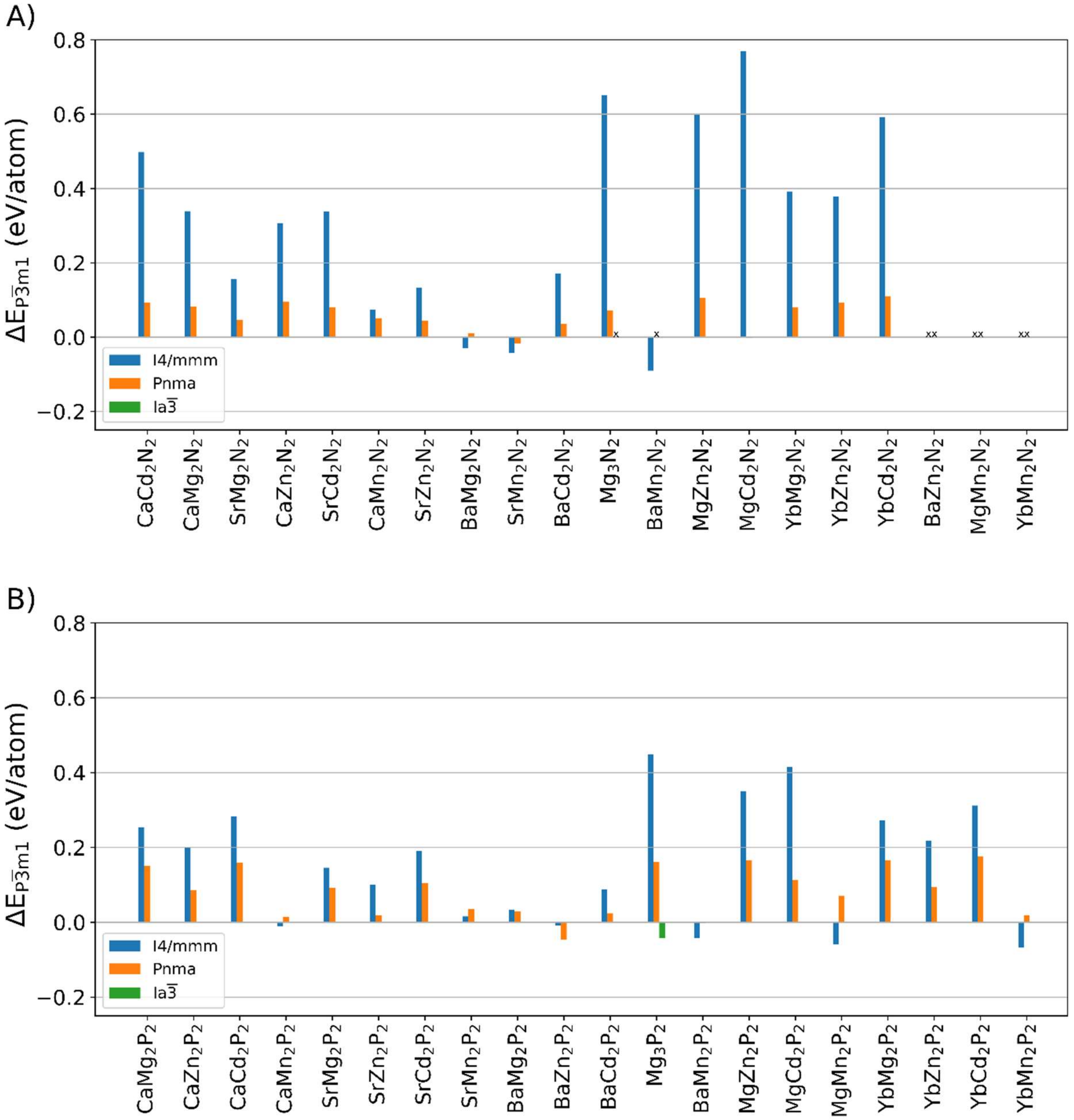

C)

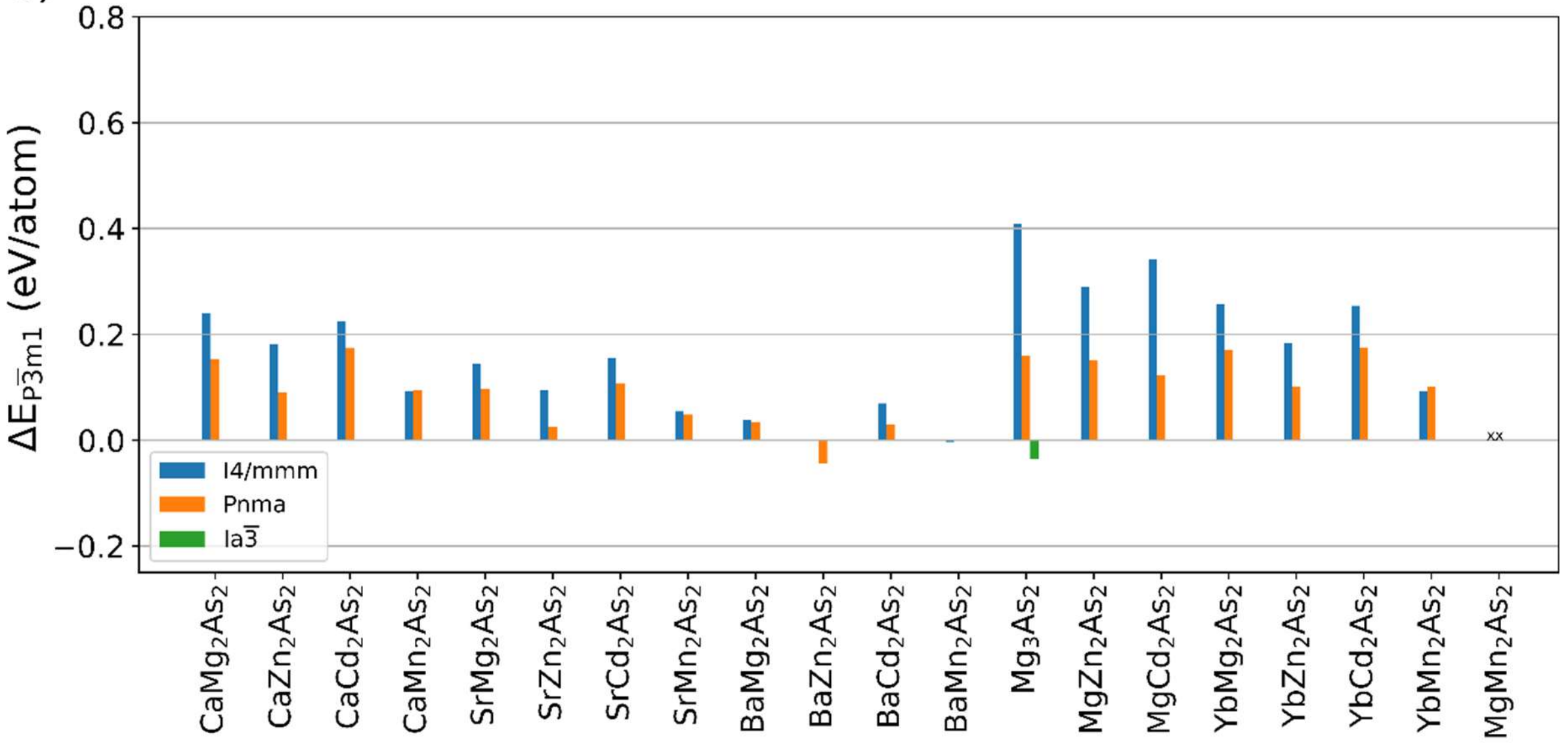

D)

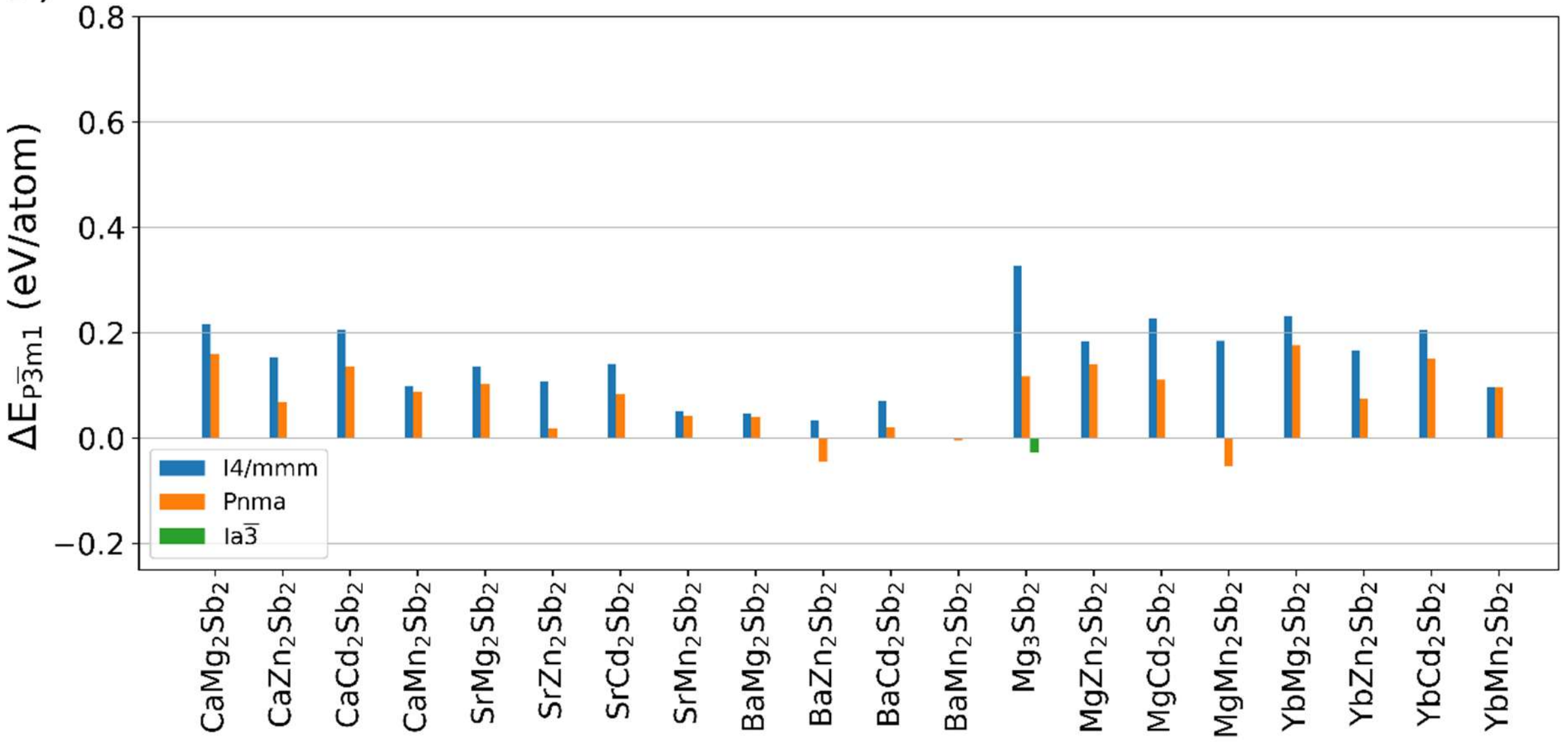

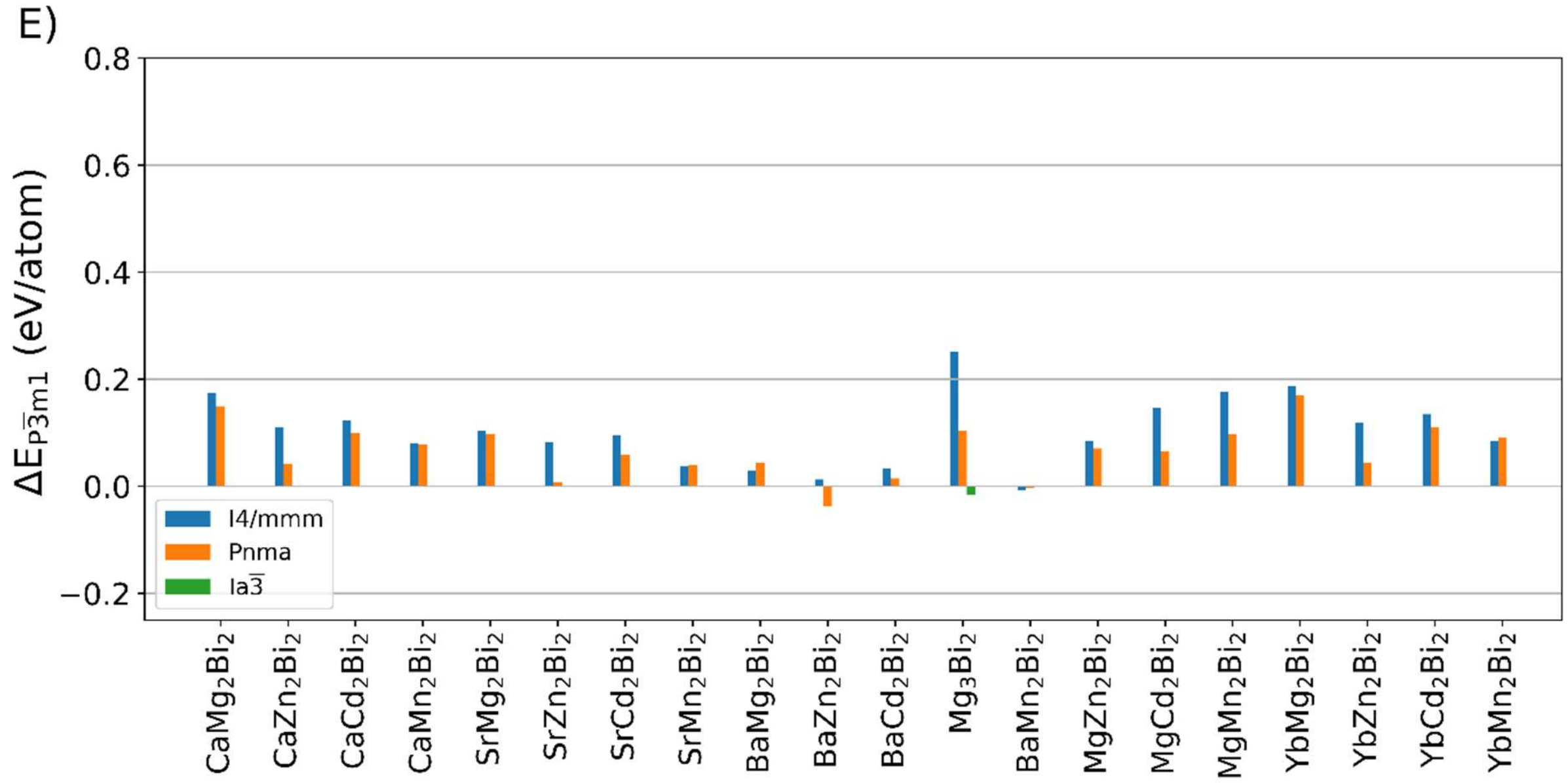

Fig. S3 – The relative stability of each $AM_2Pn_2$ compound. Here we use the final energy (eV/atom) of the $P\bar{3}m1$ polymorph as a reference to show to difference in energy of the other polymorphs. The plots are divided in A) nitrides B) phosphides C) arsenides D) antimonides and E) bismuthides

Table S1 - Stability information for $AM_2Pn_2$ compounds

| Formula | Ground State Structure | Ehull (meV/atom) | Decomposition Products |
|---|---|---|---|
| **$BaCd_2As_2$** | $P\bar{3}m1$ | -80 | 0.5 $Ba_2Cd_2As_3$ + 0.25 $Cd_3As_2$ + 0.25 Cd |
| **$BaCd_2Bi_2$** | $P\bar{3}m1$ | -21 | 0.5 $Ba_2Cd_3Bi_4$ + 0.5 Cd |
| **$BaCd_2N_2$** | $P\bar{3}m1$ | 371 | 0.182 $BaCd_{11}$ + 0.182 $N_2$ + 0.818 $BaN_2$ |
| **$BaCd_2P_2$** | $P\bar{3}m1$ | -135 | 0.364 $BaP_3$ + 0.182 $BaCd_{11}$ + 0.455 $BaP_2$ |
| **$BaCd_2Sb_2$** | $P\bar{3}m1$ | -56 | 0.5 $Ba_2Cd_2Sb_3$ + 0.5 CdSb + 0.5 Cd |
| **$BaMg_2As_2$** | $P\bar{3}m1$ | -178 | 0.042 $Ba_4As_3$ + 0.667 $Mg_3As_2$ + 0.042 $Ba_{20}As_{13}$ |
| **$BaMg_2Bi_2$** | $P\bar{3}m1$ | -186 | 0.545 $Mg_3Bi_2$ + 0.091 $Ba_{11}Bi_{10}$ + 0.364 Mg |
| **$BaMg_2N_2$** | *I4/mmm* | 51 | 0.667 $Mg_3N_2$ + 0.444 $Ba_2N$ + 0.111 $BaN_2$ |
| **$BaMg_2P_2$** | $P\bar{3}m1$ | -163 | 0.667 $Mg_3P_2$ + 0.333 $Ba_3P_2$ |
| **$BaMg_2Sb_2$** | $P\bar{3}m1$ | -181 | 0.6 $Mg_3Sb_2$ + 0.2 $Ba_5Sb_4$ + 0.2 Mg |
| **$BaMn_2As_2$** | *I4/mmm* | -169 | 0.084 $Mn_{23}As_{16}$ + 0.05 $Ba_{20}As_{13}$ + 0.059 Mn |
| **$BaMn_2Bi_2$** | *I4/mmm* | -20 | 0.5 Bi + 2 Mn + 0.5 $Ba_2Bi_3$ |
| **$BaMn_2N_2$** | *I4/mmm* | 226 | 0.3 $Mn_4N$ + 0.8 $BaMnN_2$ + 0.1 $Ba_2N$ |
| **$BaMn_2P_2$** | *I4/mmm* | -76 | 0.125 $Ba_5P_4$ + 1 $Mn_2P$ + 0.125 $Ba_3P_4$ |
| **$BaMn_2Sb_2$** | *I4/mmm* | -66 | 0.046 $Mn_{27}Sb_{26}$ + 0.2 $Ba_5Sb_4$ + 0.754 Mn |
| **$BaZn_2As_2$** | *Pnma* | -129 | 0.326 $Ba_3Zn_2As_4$ + 0.023 $BaZn_{13}$ + 0.349 $Zn_3As_2$ |
| **$BaZn_2Bi_2$** | *Pnma* | 11 | 0.154 $BaZn_{13}$ + 0.731 Bi + 0.423 $Ba_2Bi_3$ |
| **$BaZn_2N_2$** | *I4/mmm* | 94 | 0.125 $Zn_3N_2$ + 0.125 $BaZn_{13}$ + 0.875 $BaN_2$ |
| **$BaZn_2P_2$** | *Pnma* | -182 | 0.378 $Zn_3P_2$ + 0.067 $BaZn_{13}$ + 0.311 $Ba_3P_4$ |
| **$BaZn_2Sb_2$** | *Pnma* | -70 | 0.071 $BaZn_{13}$ + 0.929 $BaZnSb_2$ + 0.143 ZnSb |
| **$CaCd_2As_2$** | $P\bar{3}m1$ | -89 | 0.5 $Ca_2CdAs_2$ + 0.5 $Cd_3As_2$ |
| **$CaCd_2Bi_2$** | $P\bar{3}m1$ | -22 | 0.111 $Ca_9Cd_4Bi_9$ + 1 Bi + 1.556 Cd |
| **$CaCd_2N_2$** | $P\bar{3}m1$ | 217 | 2 Cd + 1 $CaN_2$ |
| **$CaCd_2P_2$** | $P\bar{3}m1$ | -105 | 0.5 $Ca_2CdP_2$ + 0.5 $CdP_2$ + 1 Cd |
| **$CaCd_2Sb_2$** | $P\bar{3}m1$ | -61 | 0.5 $Ca_2CdSb_2$ + 1 CdSb + 0.5 Cd |
| **$CaMg_2As_2$** | $P\bar{3}m1$ | -159 | 0.625 $Mg_3As_2$ + 0.25 $Ca_4As_3$ + 0.125 Mg |
| **$CaMg_2Bi_2$** | $P\bar{3}m1$ | -148 | 0.545 $Mg_3Bi_2$ + 0.091 $Ca_{11}Bi_{10}$ + 0.364 Mg |
| **$CaMg_2N_2$** | $P\bar{3}m1$ | -49 | 0.667 $Mg_3N_2$ + 0.333 $Ca_3N_2$ |
| **$CaMg_2P_2$** | $P\bar{3}m1$ | -219 | 0.5 $Mg_3P_2$ + 0.5 Mg + 1 CaP |
| **$CaMg_2Sb_2$** | $P\bar{3}m1$ | -135 | 0.55 $Mg_3Sb_2$ + 0.1 $Ca_{10}Mg_2Sb_9$ + 0.15 Mg |
| **$CaMn_2As_2$** | $P\bar{3}m1$ | -120 | 0.25 $Ca_4As_3$ + 0.078 $Mn_{23}As_{16}$ + 0.203 Mn |
| **$CaMn_2Bi_2$** | $P\bar{3}m1$ | -9 | 2 Mn + 1 $CaBi_2$ |
| **$CaMn_2N_2$** | $P\bar{3}m1$ | 203 | 0.167 $Mn_4N$ + 0.333 $Ca_3Mn_3N_5$ + 0.167 $Mn_2N$ |
| **$CaMn_2P_2$** | *I4/mmm* | 12 | 1 $Mn_2P$ + 1 CaP |
| **$CaMn_2Sb_2$** | $P\bar{3}m1$ | -16 | 0.071 $Ca_{14}MnSb_{11}$ + 0.047 $Mn_{27}Sb_{26}$ + 0.668 Mn |
| **$CaZn_2As_2$** | $P\bar{3}m1$ | -147 | 0.002 $Ca_4Zn_{51}$ + 0.047 $Ca_{21}Zn_4As_{18}$ + 0.574 $Zn_3As_2$ |
| **$CaZn_2Bi_2$** | $P\bar{3}m1$ | 49 | 0.039 $Ca4Zn_{51}$ + 0.314 Bi + 0.843 $CaBi_2$ |
| **$CaZn_2N_2$** | $P\bar{3}m1$ | -151 | 0.032 $Ca4Zn_{51}$ + 0.127 $Zn_3N_2$ + 0.873 $CaN_2$ |
| **$CaZn_2P_2$** | $P\bar{3}m1$ | -185 | 0.518 $Zn_3P_2$ + 0.009 $Ca_4Zn_{51}$ + 0.965 CaP |
| **$CaZn_2Sb_2$** | $P\bar{3}m1$ | -47 | 0.01 $Ca_4Zn_{51}$ + 0.106 $Ca_9Zn_4Sb_9$ + 1.042 ZnSb |
| **$Mg_3As_2$** | $Ia\bar{3}$ | -332 | 1 $Mg_2As$ + 1 MgAs |
| **$Mg_3Bi_2$** | $Ia\bar{3}$ | -187 | 0.5 $Mg_3Bi$ + 1.5 MgBi |
| **$Mg_3N_2$** | $P\bar{3}m1$ | -815 | 1 $N_2$ + 3 Mg |
| **$Mg_3P_2$** | $Ia\bar{3}$ | -362 | 1 $Mg_2P$ + 1 MgP |
| **$Mg_3Sb_2$** | $Ia\bar{3}$ | -264 | 1 $Mg_2Sb$ + 1 MgSb |
| **$MgCd_2As_2$** | $P\bar{3}m1$ | 56 | 0.333 $Mg_3As_2$ + 0.667 $Cd_3As_2$ |
| **$MgCd_2Bi_2$** | $P\bar{3}m1$ | 48 | 0.333 $Mg_3Bi_2$ + 1.333 Bi + 2 Cd |
| **$MgCd_2N_2$** | *Pnma* | 440 | 0.667 $N_2$ + 2 Cd + 0.333 $Mg_3N_2$ |
| **$MgCd_2P_2$** | $P\bar{3}m1$ | 59 | 0.333 $Mg_3P_2$ + 0.667 $CdP_2$ + 1.333 Cd |
| **$MgCd_2Sb_2$** | $P\bar{3}m1$ | 40 | 0.333 $Mg_3Sb_2$ + 1.333 CdSb + 0.667 Cd |
| **$MgMn_2As_2$** | *Pnma* | 179 | 0.333 $Mg_3As_2$ + 0.083 $Mn_{23}As_{16}$ + 0.083 Mn |
| **$MgMn_2Bi_2$** | $P\bar{3}m1$ | 93 | 0.333 $Mg_3Bi_2$ + 1.333 Bi + 2 Mn |
| **$MgMn_2N_2$** | *I4/mmm* | 125 | 1 $Mn_2N$ + 0.167 $N_2$ + 0.333 $Mg_3N_2$ |
| **$MgMn_2P_2$** | *I4/mmm* | 96 | 0.667 Mn2P + 0.333 Mg3P2 + 0.667 MnP |
| **$MgMn_2Sb_2$** | *Pnma* | 47 | 0.333 $Mg_3Sb_2$ + 0.051 $Mn_{27}Sb_{26}$ + 0.615 Mn |
| **$MgZn_2As_2$** | $P\bar{3}m1$ | -7 | 0.333 $Mg_3As_2$ + 0.667 $Zn_3As_2$ |
| **$MgZn_2Bi_2$** | $P\bar{3}m1$ | 121 | 0.333 $Mg_3Bi_2$ + 1.333 Bi + 2 Zn |
| **$MgZn_2N_2$** | $P\bar{3}m1$ | 5 | 0.667 $Zn_3N_2$ + 0.333 $Mg_3N_2$ |
| **$MgZn_2P_2$** | $P\bar{3}m1$ | -18 | 0.667 $Zn_3P_2$ + 0.333 $Mg_3P_2$ |
| **$MgZn_2Sb_2$** | $P\bar{3}m1$ | 32 | 0.333 $Mg_3Sb_2$ + 1.333 ZnSb + 0.667 Zn |
| **$SrCd_2As_2$** | $P\bar{3}m1$ | -115 | 0.5 $Sr_2CdAs_2$ + 0.5 $Cd_3As_2$ |
| **$SrCd_2Bi_2$** | $P\bar{3}m1$ | -24 | 1 $SrCdBi_2$ + 1 Cd |
| **$SrCd_2N_2$** | $P\bar{3}m1$ | 258 | 0.182 $SrCd_{11}$ + 0.818 $SrN_2$ + 0.182 $N_2$ |
| **$SrCd_2P_2$** | *$P\bar{3}m1$* | -163 | 0.333 $Sr_3P_4$ + 0.333 $CdP_2$ + 1.667 Cd |
| **$SrCd_2Sb_2$** | *$P\bar{3}m1$* | -140 | 0.5 CdSb + 1.5 Cd + 0.5 $Sr_2Sb_3$ |
| **$SrMg_2As_2$** | *$P\bar{3}m1$* | -189 | 0.625 $Mg_3As_2$ + 0.25 $Sr_4As_3$ + 0.125 Mg |
| **$SrMg_2Bi_2$** | *$P\bar{3}m1$* | -174 | 0.545 $Mg_3Bi_2$ + 0.091 $Sr_{11}Bi_{10}$ + 0.364 Mg |
| **$SrMg_2N_2$** | *$P\bar{3}m1$* | -32 | 0.333 $Sr_2N$ + 0.667 $Mg_3N_2$ + 0.333 SrN |
| **$SrMg_2P_2$** | *$P\bar{3}m1$* | -183 | 0.667 $Mg_3P_2$ + 0.333 $Sr_3P_2$ |
| **$SrMg_2Sb_2$** | *$P\bar{3}m1$* | -190 | 0.667 $Mg_3Sb_2$ + 0.02 $Sr_{11}Sb_{10}$ + 0.157 $Sr_5Sb_3$ |
| **$SrMn_2As_2$** | *$P\bar{3}m1$* | -147 | 0.25 $Sr_4As_3$ + 0.078 $Mn_{23}As_{16}$ + 0.203 Mn |
| **$SrMn_2Bi_2$** | *$P\bar{3}m1$* | -10 | 0.5 Bi + 2 Mn + 0.5 $Sr_2Bi_3$ |
| **$SrMn_2N_2$** | *I4/mmm* | 216 | 0.3 $Mn_4N$ + 0.1 $Sr_2N$ + 0.8 $SrMnN_2$ |
| **$SrMn_2P_2$** | *$P\bar{3}m1$* | -44 | 1 $Mn_2P$ + 1 SrP |
| **$SrMn_2Sb_2$** | *$P\bar{3}m1$* | -63 | 0.091 $Sr_{11}Sb_{10}$ + 0.042 $Mn_{27}Sb_{26}$ + 0.867 Mn |
| **$SrZn_2As_2$** | *$P\bar{3}m1$* | -83 | 0.018 $SrZn_{13}$ + 0.491 $Sr_2Zn_2As_3$ + 0.263 $Zn_3As_2$ |
| **$SrZn_2Bi_2$** | *$P\bar{3}m1$* | 50 | 0.154 $SrZn_{13}$ + 0.731 Bi + 0.423 $Sr_2Bi_3$ |
| **$SrZn_2N_2$** | *$P\bar{3}m1$* | -34 | 0.154 $SrZn_{13}$ + 0.846 $SrN_2$ + 0.154 $N_2$ |
| **$SrZn_2P_2$** | *$P\bar{3}m1$* | -185 | 0.034 $SrZn_{13}$ + 0.517 $Zn_3P_2$ + 0.966 SrP |
| **$SrZn_2Sb_2$** | *$P\bar{3}m1$* | -39 | 0.929 $SrZn_5b_2$ + 0.071 $SrZn_{13}$ + 0.143 ZnSb |
| **$YbCd_2As_2$** | *$P\bar{3}m1$* | -169 | 0.625 $Cd_3As_2$ + 0.125 Cd + 0.25 $Yb_4As_3$ |
| **$YbCd_2Bi_2$** | *$P\bar{3}m1$* | -43 | 2 Cd + 1 $YbBi_2$ |
| **$YbCd_2N_2$** | *$P\bar{3}m1$* | 183 | 2 Cd + 1 $YbN_2$ |
| **$YbCd_2P_2$** | *$P\bar{3}m1$* | -314 | 0.333 $YbCd_6$ + 0.333 $YbP_5$ + 0.333 YbP |
| **$YbCd_2Sb_2$** | *$P\bar{3}m1$* | -61 | 0.5 $Yb_2CdSb_2$ + 1 CdSb + 0.5 Cd |
| **$YbMg_2As_2$** | *$P\bar{3}m1$* | -171 | 0.625 $Mg_3As_2$ + 0.125 Mg + 0.25 $Yb_4As_3$ |
| **$YbMg_2Bi_2$** | *$P\bar{3}m1$* | -133 | 0.625 $Mg_3Bi_2$ + 0.25 $Yb_4Bi_3$ + 0.125 Mg |
| **$YbMg_2N_2$** | *$P\bar{3}m1$* | -82 | 0.667 $Mg_3N_2$ + 0.333 $Yb_3N_2$ |
| **$YbMg_2P_2$** | *$P\bar{3}m1$* | -371 | 0.667 $Mg_3P_2$ + 0.167 $Yb_3P$ + 0.5 YbP |
| **$YbMg_2Sb_2$** | *$P\bar{3}m1$* | -148 | 0.625 $Mg_3Sb_2$ + 0.25 $Yb_4Sb_3$ + 0.125 Mg |
| **$YbMn_2As_2$** | *$P\bar{3}m1$* | -144 | 0.078 $Mn_{23}As_{16}$ + 0.203 Mn + 0.25 $Yb_4As_3$ |
| **$YbMn_2Bi_2$** | *$P\bar{3}m1$* | -3 | 2 Mn + 1 $YbBi_2$ |
| **$YbMn_2N_2$** | *Pnma* | 16 | 1 $Mn_2N$ + 0.25 $Yb_3N_2$ + 0.25 $YbN_2$ |
| **$YbMn_2P_2$** | *I4/mmm* | -173 | 0.286 MnP + 0.143 $Yb_2Mn_{12}P_7$ + 0.714 YbP |
| **$YbMn_2Sb_2$** | *$P\bar{3}m1$* | -44 | 0.048 $Mn_{27}Sb_{26}$ + 0.25 $Yb_4Sb_3$ + 0.702 Mn |
| **$YbZn_2As_2$** | *$P\bar{3}m1$* | -213 | 0.01 $YbZn_{11}$ + 0.629 $Zn_3As_2$ + 0.247 $Yb_4As_3$ |
| **$YbZn_2Bi_2$** | *$P\bar{3}m1$* | 55 | 0.182 $YbZn_{11}$ + 0.364 Bi + 0.818 $YbBi_2$ |
| **$YbZn_2N_2$** | *$P\bar{3}m1$* | -231 | 0.143 $YbZn_{11}$ + 0.143 $Zn_3N_2$ + 0.857 $YbN_2$ |
| **$YbZn_2P_2$** | *$P\bar{3}m1$* | -387 | 0.04 $YbZn_{11}$ + 0.52 $Zn_3P_2$ + 0.96 YbP |
| **$YbZn_2Sb_2$** | *$P\bar{3}m1$* | -48 | 0.478 $Yb_2ZnSb_2$ + 0.043 $YbZn_{11}$ + 1.043 ZnSb |

Table S2: Electronic Structure data for P-3m1 $AM_2Pn_2$ Compounds

| Formula | Indirect Bandgap (eV) | Direct Bandgap (eV) | Bandgap Type (D: Direct, I: Indirect) |
|---|---|---|---|
| **$BaCd_2As_2$** | 0.68 | 0.68 | D |
| **$BaCd_2Bi_2$** | 0 | 0 | D |
| **$BaCd_2N_2$** | 0.54 | 0.57 | I |
| **$BaCd_2P_2$** | 1.31 | 1.31 | D |
| **$BaCd_2Sb_2$** | 0.32 | 0.34 | D |
| **$BaMg_2As_2$** | 1.61 | 1.9 | I |
| **$BaMg_2Bi_2$** | 0.49 | 0.49 | D |
| **$BaMg_2P_2$** | 1.75 | 2.46 | I |
| **$BaMg_2Sb_2$** | 1.28 | 1.86 | I |
| **$BaZn_2As_2$** | 0.67 | 0.77 | I |
| **$BaZn_2Bi_2$** | 0 | 0 | D |
| **$BaZn_2P_2$** | 1.1 | 1.44 | I |
| **$BaZn_2Sb_2$** | 0 | 0 | D |
| **$BaMg_2N_2$** | 1.89 | 2.24 | I |
| **$CaCd_2As_2$** | 0.81 | 0.81 | D |
| **$CaCd_2Bi_2$** | 0 | 0 | D |
| **$CaCd_2N_2$** | 0.47 | 0.54 | I |
| **$CaCd_2P_2$** | 1.5 | 1.5 | D |
| **$CaCd_2Sb_2$** | 0.5 | 0.57 | I |
| **$CaMg_2As_2$** | 2 | 2.2 | I |
| **$CaMg_2Bi_2$** | 0.39 | 0.39 | D |
| **$CaMg_2P_2$** | 2.24 | 3.13 | I |
| **$CaMg_2Sb_2$** | 1.37 | 1.88 | I |
| **$CaZn_2As_2$** | 1.02 | 1.02 | D |
| **$CaZn_2Bi_2$** | 0 | 0 | D |
| **$CaZn_2N_2$** | 1.64 | 1.64 | D |
| **$CaZn_2P_2$** | 1.58 | 1.83 | I |
| **$CaZn_2Sb_2$** | 0.12 | 0.14 | D |
| **$CaMg_2N_2$** | 3.03 | 3.03 | D |
| **$MgCd_2As_2$** | 0.4 | 0.45 | I |
| **$MgCd_2Bi_2$** | 0 | 0 | D |
| **$MgCd_2N_2$** | 0.15 | 0.41 | I |
| **$MgCd_2P_2$** | 1.04 | 1.12 | I |
| **$MgCd_2Sb_2$** | 0 | 0 | D |
| **$MgMn_2Bi_2$** | 0 | 0 | D |
| **$MgMn_2P_2$** | 0 | 0 | D |
| **$MgMn_2Sb_2$** | 0.24 | 1.64 | I |
| **$MgZn_2As_2$** | 0.57 | 1.24 | I |
| **$MgZn_2Bi_2$** | 0 | 0 | D |
| **$MgZn_2N_2$** | 1.3 | 1.3 | D |
| **$MgZn_2P_2$** | 1.38 | 1.94 | I |
| **$MgZn_2Sb_2$** | 0 | 0 | D |
| **$Mg_3As_2$** | 1.49 | 1.58 | I |
| **$Mg_3Bi_2$** | 0 | 0 | D |
| **$Mg_3N_2$** | 2.64 | 2.65 | D |
| **$Mg_3P_2$** | 1.81 | 2.53 | I |
| **$Mg_3Sb_2$** | 0.55 | 1.54 | I |
| **$SrCd_2As_2$** | 0.71 | 0.71 | D |
| **$SrCd_2Bi_2$** | 0 | 0 | D |
| **$SrCd_2N_2$** | 0.48 | 0.53 | I |
| **$SrCd_2P_2$** | 1.35 | 1.35 | D |
| **$SrCd_2Sb_2$** | 0.34 | 0.4 | I |
| **$SrMg_2As_2$** | 1.94 | 2.1 | I |
| **$SrMg_2Bi_2$** | 0.37 | 0.37 | D |
| **$SrMg_2P_2$** | 2.14 | 2.86 | I |
| **$SrMg_2Sb_2$** | 1.43 | 1.92 | I |
| **$SrZn_2As_2$** | 0.88 | 0.88 | D |
| **$SrZn_2Bi_2$** | 0 | 0 | D |
| **$SrZn_2N_2$** | 1.41 | 1.41 | D |
| **$SrZn_2P_2$** | 1.52 | 1.66 | I |
| **$SrZn_2Sb_2$** | 0.09 | 0.13 | I |
| **$SrMg_2N_2$** | 2.64 | 2.64 | D |
| **$YbCd_2As_2$** | 0.75 | 0.75 | D |
| **$YbCd_2Bi_2$** | 0 | 0 | D |
| **$YbCd_2N_2$** | 0.42 | 0.51 | I |
| **$YbCd_2P_2$** | 1.44 | 1.44 | D |
| **$YbCd_2Sb_2$** | 0.48 | 0.58 | I |
| **$YbMg_2As_2$** | 1.86 | 2.12 | I |
| **$YbMg_2Bi_2$** | 0.33 | 0.33 | D |
| **$YbMg_2P_2$** | 2.09 | 3.03 | I |
| **$YbMg_2Sb_2$** | 1.21 | 1.84 | I |
| **$YbZn_2As_2$** | 1.09 | 1.1 | D |
| **$YbZn_2Bi_2$** | 0 | 0 | D |
| **$YbZn_2N_2$** | 1.65 | 1.65 | D |
| **$YbZn_2P_2$** | 1.45 | 1.83 | I |
| **$YbZn_2Sb_2$** | 0.11 | 0.15 | I |
| **$YbMg_2N_2$** | 2.94 | 2.94 | D |

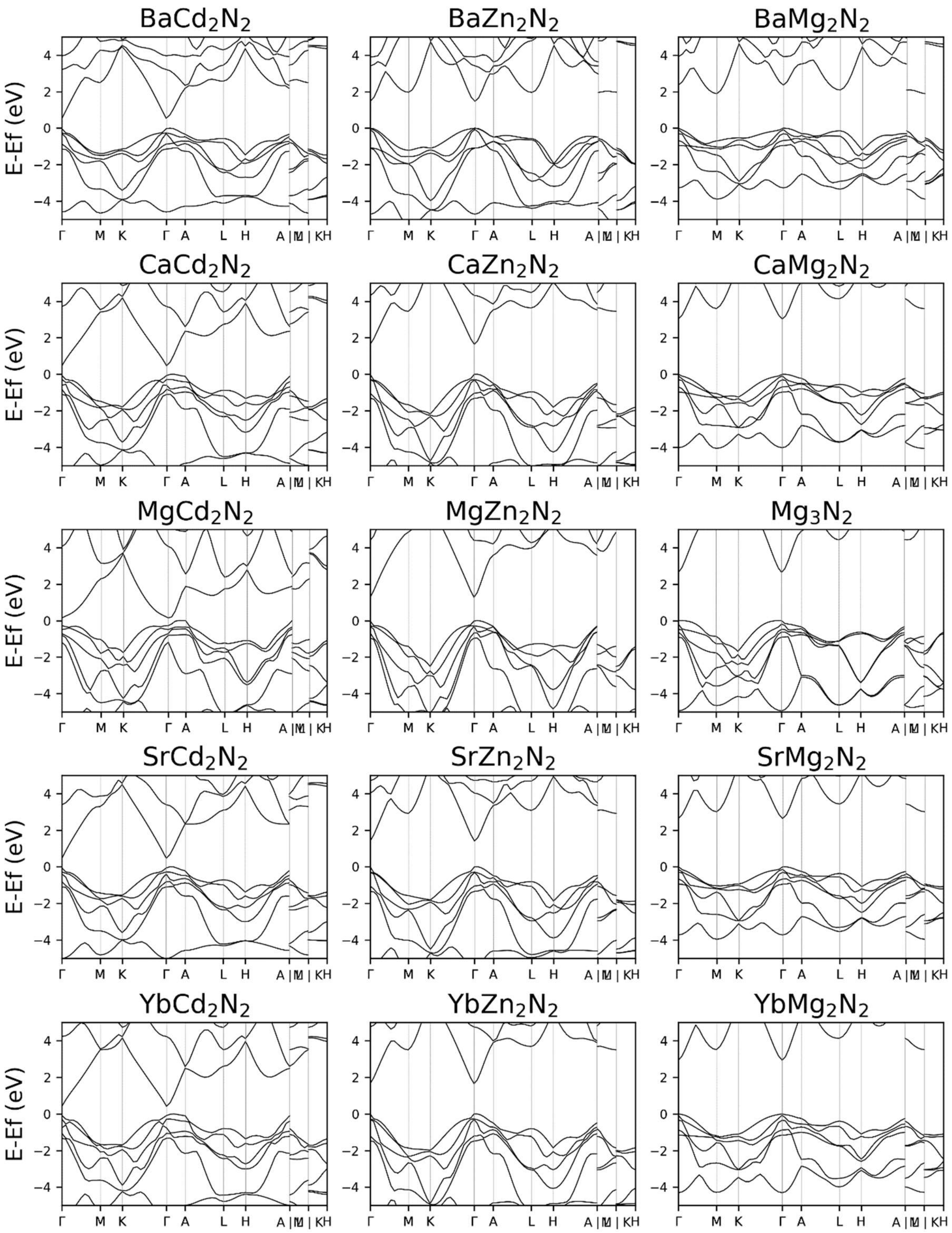
$BaCd_2N_2$
$BaZn_2N_2$
$BaMg_2N_2$
$CaCd_2N_2$
$CaZn_2N_2$
$CaMg_2N_2$
$MgCd_2N_2$
$MgZn_2N_2$
$Mg_3N_2$
$SrCd_2N_2$
$SrZn_2N_2$
$SrMg_2N_2$
$YbCd_2N_2$
$YbZn_2N_2$
$YbMg_2N_2$
E-Ef (eV)
Γ M K Γ A L H A|M L|K H

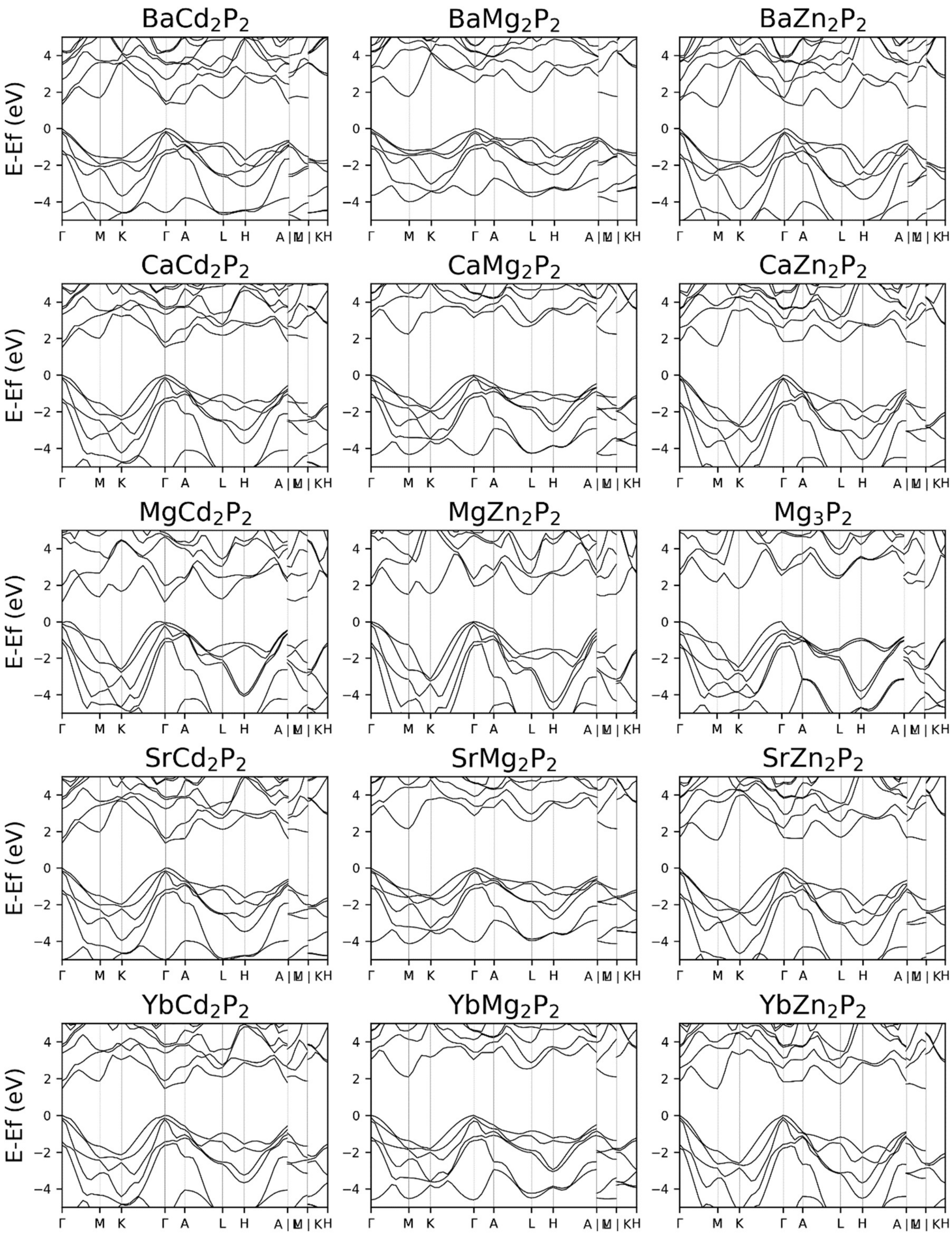

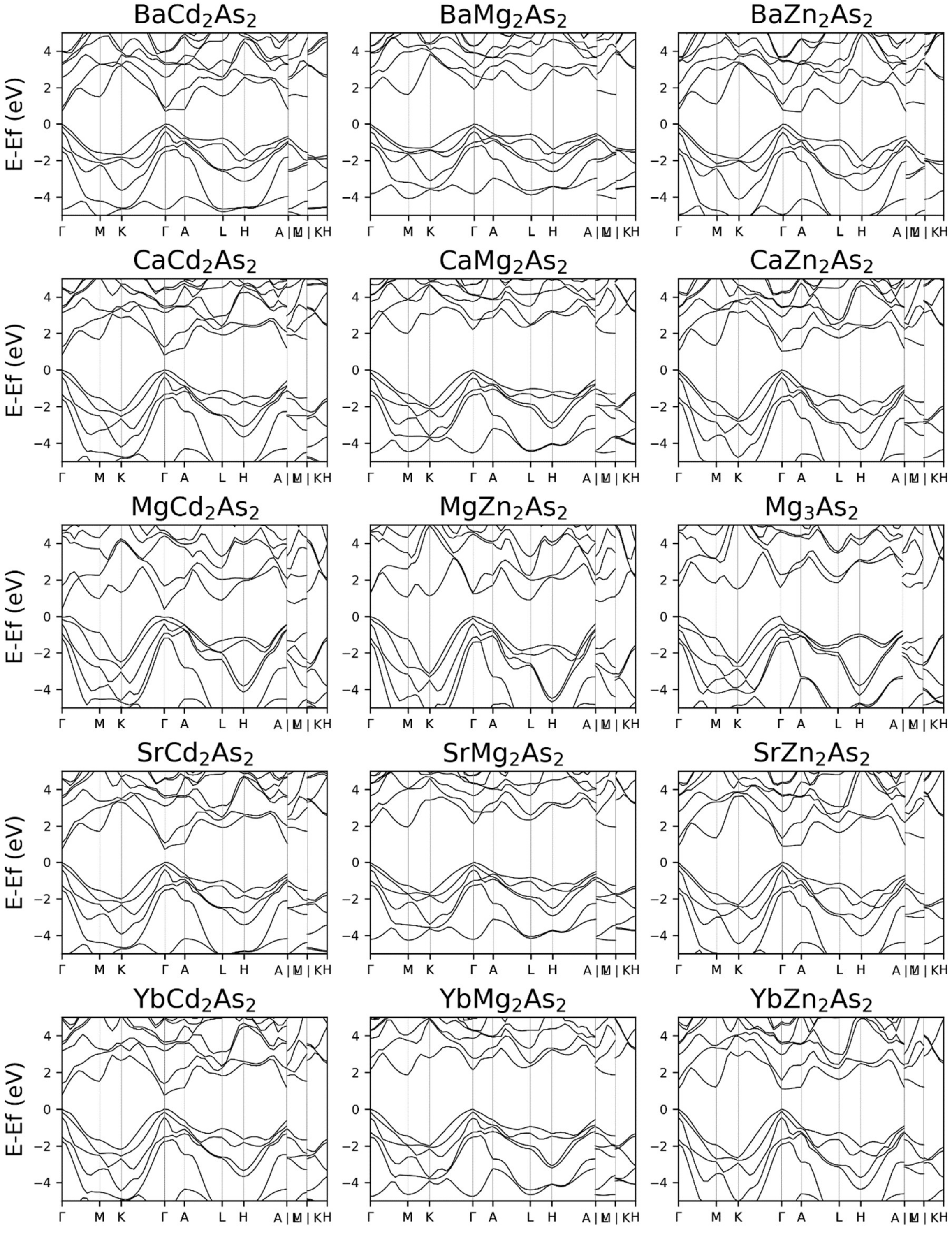
$BaCd_2As_2$
$BaMg_2As_2$
$BaZn_2As_2$
$CaCd_2As_2$
$CaMg_2As_2$
$CaZn_2As_2$
$MgCd_2As_2$
$MgZn_2As_2$
$Mg_3As_2$
$SrCd_2As_2$
$SrMg_2As_2$
$SrZn_2As_2$
$YbCd_2As_2$
$YbMg_2As_2$
$YbZn_2As_2$
E-Ef (eV)
4
2
0
−2
−4
Γ M K Γ A L H A|L M|K H

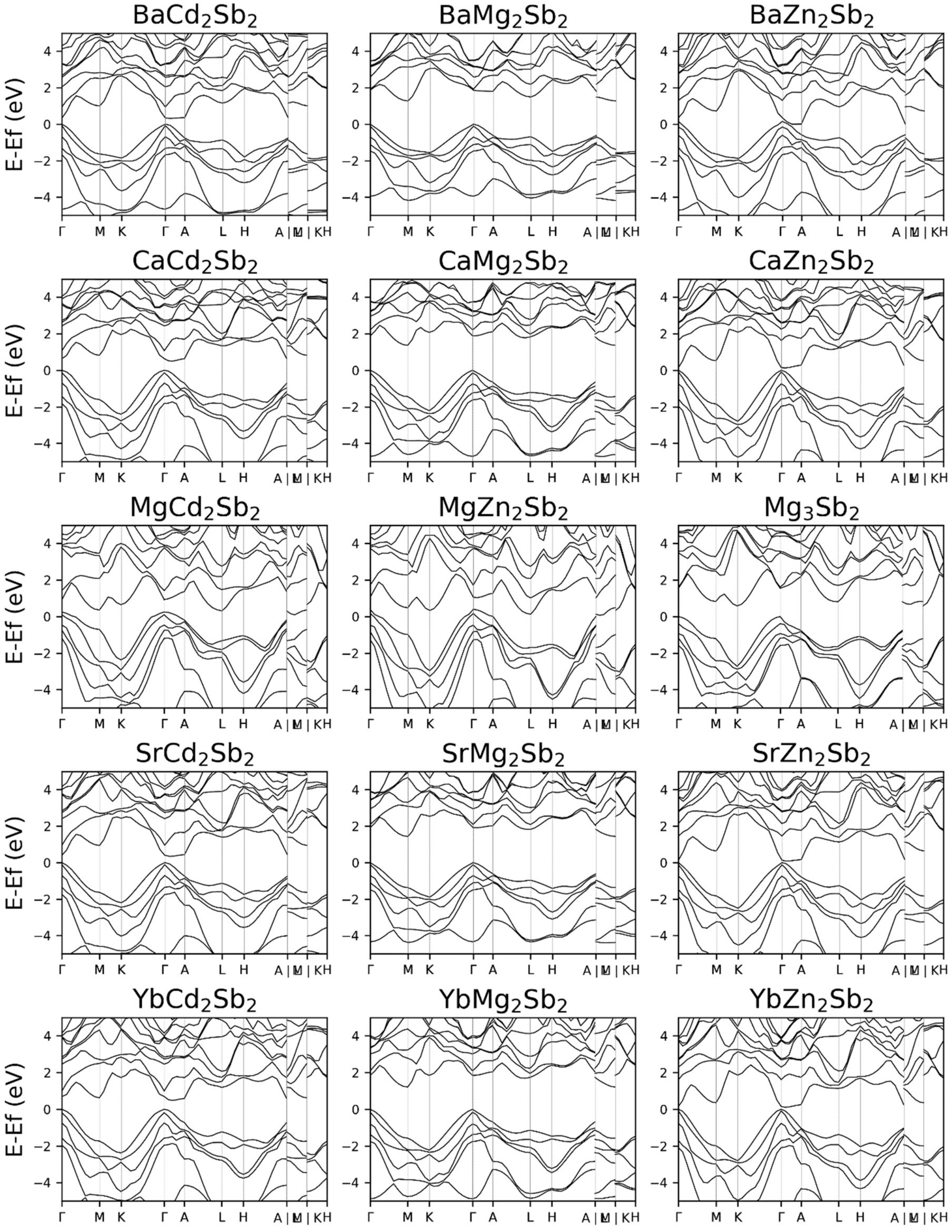
BaCd2Sb2
BaMg2Sb2
BaZn2Sb2
CaCd2Sb2
CaMg2Sb2
CaZn2Sb2
MgCd2Sb2
MgZn2Sb2
Mg3Sb2
SrCd2Sb2
SrMg2Sb2
SrZn2Sb2
YbCd2Sb2
YbMg2Sb2
YbZn2Sb2
E-Ef (eV)
4
2
0
−2
−4
Γ M K Γ A L H A|L M|K H

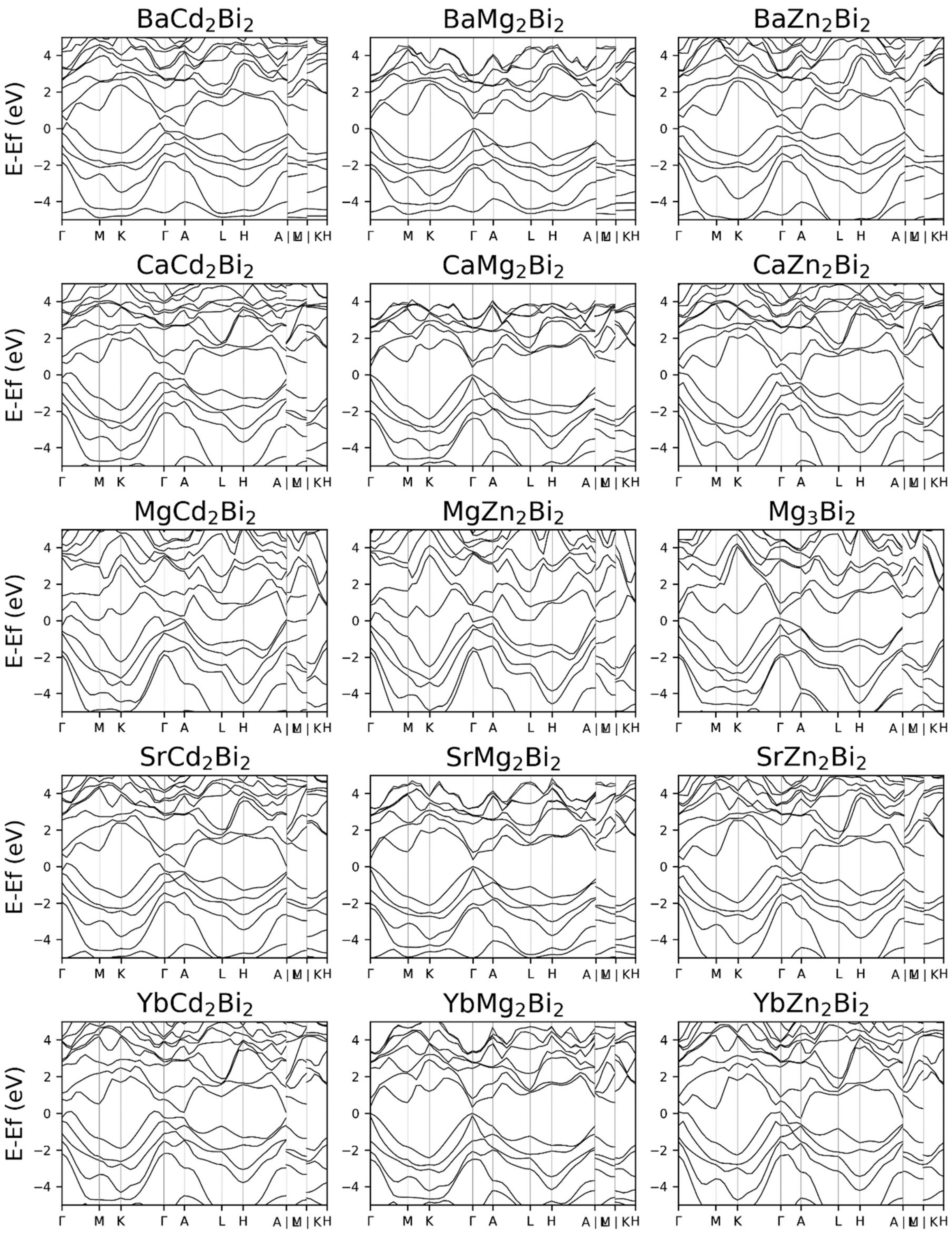

Fig. S4 – HSE+SOC band structures of $AM_2Pn_2$ compounds in the P$\bar{3}$m1 space group structure.

*SI Section 2: Comparison of Experimentally Reported Bandgaps and HSE+SOC on PBE Calculated Bandgaps*

We find several discrepancies between our calculated bandgaps and those reported in the literature that we have chosen to exclude from the MAE of our comparison, which we will discuss now. First, several materials have their bandgaps reported several times independently, so we can compare between those. Ponnambalam et. al [2] report a bandgap of 0.6 eV for $CaZn_2P_2$ deduced from temperature dependence of electrical resistivity in powder/crystal samples, whereas our calculations show a much larger indirect bandgap of 1.58 eV. Further experimental reports also claim a higher bandgap in agreement with our calculations; 1.85 eV in Katsube et al. [3] by diffuse reflectance and 1.6 eV in Quadir et al. [4] by photoluminescence spectroscopy, leading us to conclude Ponnambalam et al. [2] misrepresented the bandgap of $CaZn_2P_2$. They also report the bandgap of YbZn2P2 to be 0.4 eV in the same publication, whereas our calculations show an indirect bandgap of 1.45 eV. To the best of our knowledge there are no other experimental reports for the bandgap of this material, but given the previous inaccuracy of their $CaZn_2P_2$ bandgap measurement, we believe this report is likely incorrect as well. Since the bandgaps of this class of materials are not very well explored, repeated bandgap measurements are somewhat sparse, so next we look at compounds that have only one reported bandgap measurement, but have similar compounds with reported bandgap measurements. Chen et al. [5] report the bandgap of $SrCd_2As_2$ is 0.21 eV via the change in electrical resistivity with respect to temperature, whereas we have calculated it to be 0.71 eV (nearly direct). While there do not seem to be other reports of the bandgap of $SrCd_2As_2$, $BaCd_2As_2$ has been reported to have a bandgap of 0.4 eV with and 0.84 eV by Kunioka et al. [6] and Yang et al. [7] , respectively, both also via electrical resistivity, and $SrCd_2Sb_2$ has been reported as 0.63 by Jin et al. [8] from a Tauc plot of optical absorbance. Based off the trend in atomic mass, $SrCd_2As_2$ should have a bandgap higher than $BaCd_2As_2$ or $SrCd_2Sb_2$ since it is lighter, corroborating our calculations. Similarly, Wang et al.'s [9] 0.7 eV measurement of the bandgap of $BaCd_2Sb_2$ does not match with our calculations (0.32 eV, direct) or reports of similar compounds: $BaCd_2As_2$'s bandgap of 0.4 and 0.84 eV by Kunioka et al. [6] and Yang et al. [7] and $SrCd_2Sb_2$'s bandgap of 0.63 eV measured by Jin et al. [8]. So, we believe 0.7 eV would be too high to realistically be the $BaCd_2Sb_2$'s bandgap since it should be lower than $BaCd_2As_2$ and $SrCd_2Sb_2$. For $CaZn_2N_2$ it appears our calculations estimate the bandgap to be too low, 1.64 eV nearly direct, versus experimental reports of 1.9 and 1.93 eV from Hinuma et al. [10] and Tsuji et al. [11] respectively. HSE band structure calculations based off of a prior HSE relaxation show a bandgap of 1.90, which matches the experimental value very well.

The reported band gaps for $CaMg_2N_2$ in the literature are quite scattered. Based on Tauc plot method, Ma et Al. [12] reported a band gap of 2.56 eV, while Hinuma et Al. [10] found a band gap of 3.25 eV. Our HSE06 calculated bandgap is direct and has an energy of 3.0 eV. Given the underestimation noted for our calculations for $CaZn_2N_2$, it is likely our HSE06+PBE calculation for the bandgap of $CaMg_2N_2$ is underestimated as well, so we do an HSE relaxation and HSE+SOC band structure calculation. This more advanced calculation gives a bandgap of 3.25 eV, matching the results of Hinuma et al. [10].As another check, the HSE+SOC band structure calculation on the PBEsol relaxed structure gives a bandgap of 3.26 eV. So, noting a similar error for $CaZn_2N_2$ it seems that our data has a slight underestimation of the bandgaps for some nitrides due to overestimation of the lattice constants by PBE, where HSE would have been better for the relaxation, or PBEsol at a similar computational cost to PBE. The only other reported nitride bandgaps in this class are $SrZn_2N_2$ by Kikuchi et al. [13] (error of 0.2 eV) and $Mg_3N_2$ [14] by Fang et al. and Ma et al. [12] (error of 0.15 and -0.06 eV, respectively), which agree decently well. Though without more reports of the bandgaps of $AM_2N_2$ compounds it is difficult to say how pervasive this trend is.

Table S3 - Literature comparison for bandgaps of $AM_2Pn_2$ compounds

| Formula | Bandgap (HSE+SOC minimum) | Bandgap (experimentally reported) | Difference | Citation |
|---|---|---|---|---|
| **$BaCd_2As_2$** | 0.68 | 0.4 | -0.28 | 6 |
| **$BaCd_2As_2$** | 0.68 | 0.84 | -0.26 | 7 |
| **$BaCd_2P_2$** | 1.31 | 1.46 | 0.15 | 15 |
| **$BaCd_2Sb_2$** | 0.32 | 0.7 | 0.38 | 9 |
| **$CaMg_2Bi_2$** | 0.40 | 0.2 | -0.19 | 16 |
| **$CaMg_2N_2$** | 3.03 | 2.56 | -0.47 | 12 |
| **$CaZn_2N_2$** | 1.64 | 1.93 | 0.29 | 11 |
| **$CaZn_2N_2$** | 1.64 | 1.9 | 0.26 | 10 |
| **$CaZn_2P_2$** | 1.58 | 0.6 | -0.98 | 2 |
| **$CaZn_2P_2$** | 1.58 | 1.85 | 0.27 | 3 |
| **$CaZn_2P_2$** | 1.58 | 1.6 | 0.02 | 4 |
| **$CaZn_2Sb_2$** | 0.12 | 0.26 | 0.14 | 17 |
| **$CaZn_2Sb_2$** | 0.12 | 0.26 | 0.14 | 18 |
| **$Mg_3Bi_2$** | 0 | 0 | 0 | 19 |
| **$Mg_3N_2$** | 2.64 | 2.8 | 0.16 | 14 |
| **$Mg_3N_2$** | 2.64 | 2.58 | -0.06 | 12 |
| **$Mg_3Sb_2$** | 0.55 | 0.54 | -0.01 | 19 |
| **$SrCd_2As_2$** | 0.71 | 0.21 | -0.50 | 5 |
| **$SrCd_2Sb_2$** | 0.34 | 0.63 | 0.29 | 8 |
| **$SrZn_2N_2$** | 1.41 | 1.6 | 0.19 | 13 |
| **$SrZn_2P_2$** | 1.52 | 1.7 | 0.18 | 3 |
| **$SrZn_2Sb_2$** | 0.09 | 0.27 | 0.18 | 17 |
| **$YbCd_2Sb_2$** | 0.48 | 0.55 | 0.07 | 20 |
| **$YbMg_2Bi_2$** | 0.33 | 0.3 | -0.03 | 16 |
| **$YbZn_2P_2$** | 1.45 | 0.4 | -1.05 | 2 |
| **$YbZn_2Sb_2$** | 0.11 | 0 | -0.11 | 17 |
| **$YbZn_2Sb_2$** | 0.11 | 0.55 | 0.44 | 20 |
| **$YbZn_2Sb_2$** | 0.11 | 0 | -0.11 | 18 |

A)

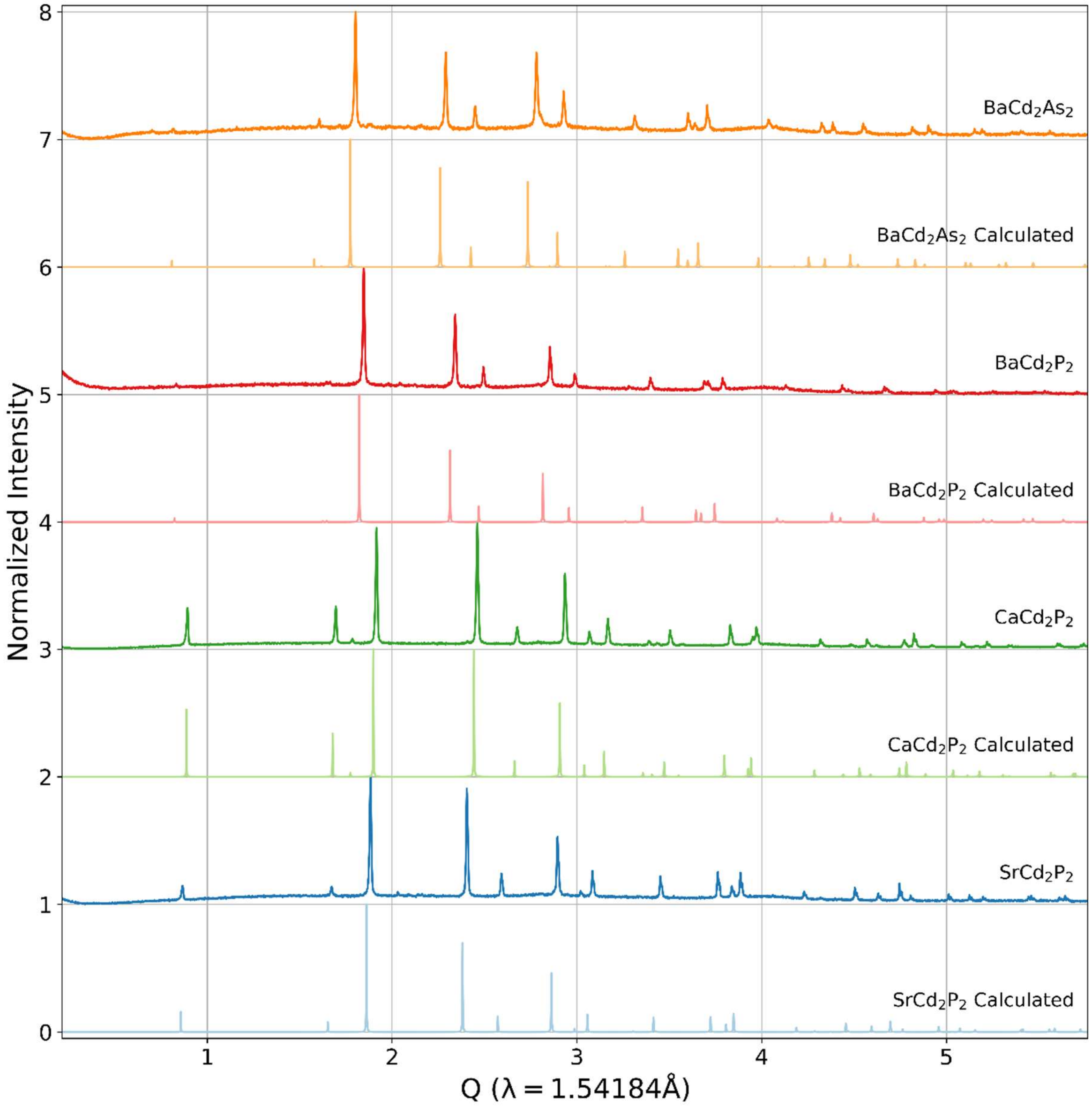

B)

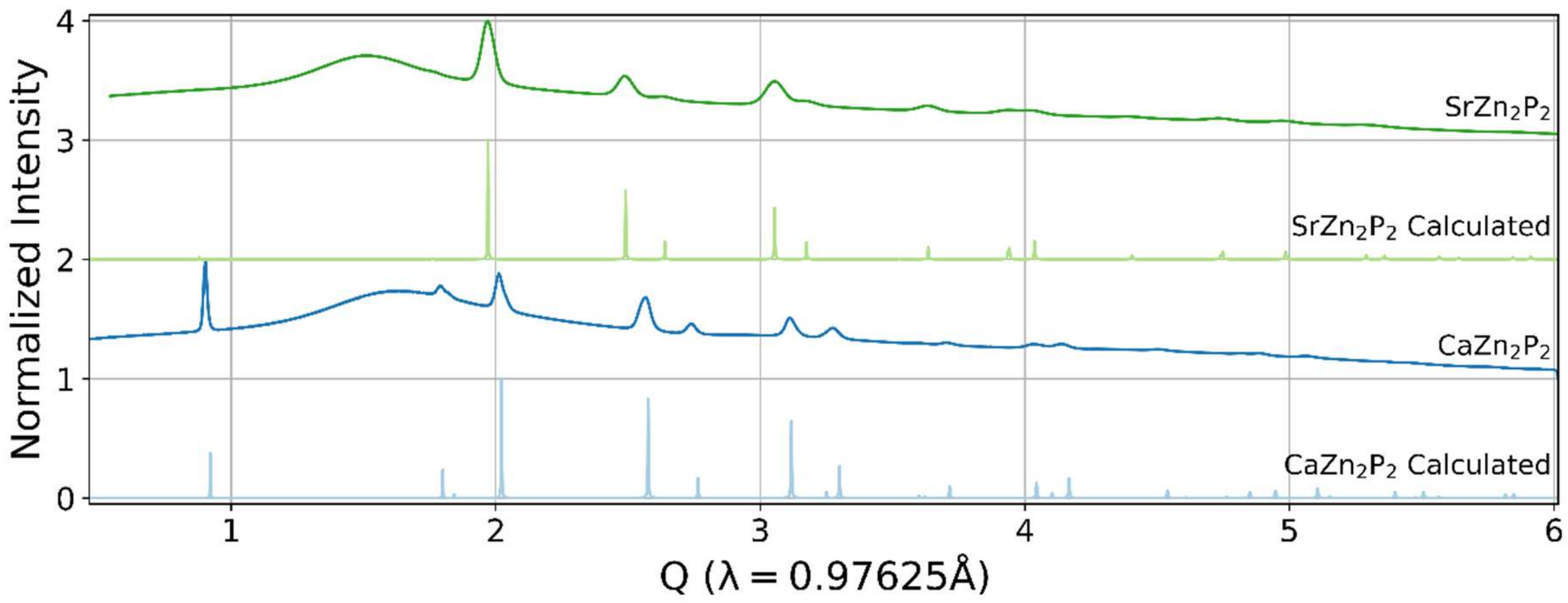

Fig. S5 – X-ray diffraction patterns (XRD) of synthesized phases in A) powder B) thin films. Calculated patterns were generated in VESTA[21] from PBE-relaxed $P\overline{3}m1$ unit cells. Note the amorphous background in the thin film samples is attributable to the amorphous silica substrate.